\documentclass[aps,prb,reprint]{revtex4-1}
\usepackage{H1}
\usepackage{amsmath}
\usepackage[pdftex,colorlinks=true]{hyperref}
\usepackage{amssymb}
\usepackage{amsthm}
\usepackage{amsfonts}
\usepackage{bbm}
\usepackage{array}
\newcolumntype{L}[1]{>{\raggedright\let\newline\\\arraybackslash\hspace{0pt}}m{#1}}
\newcolumntype{C}[1]{>{\centering\let\newline\\\arraybackslash\hspace{0pt}}m{#1}}
\newcolumntype{R}[1]{>{\raggedleft\let\newline\\\arraybackslash\hspace{0pt}}m{#1}}

\newcolumntype{N}{@{}m{0pt}@{}}

\begin{document}
\title{ Absolute Stability and Spatiotemporal Long-Range Order in Floquet systems}
\author{C.~W.~von~Keyserlingk}
\thanks{These authors contributed equally to the preparation of this work.}
\affiliation{Department of Physics, Princeton University, Princeton, New Jersey 08544, USA}
\author{Vedika~Khemani}
\thanks{These authors contributed equally to the preparation of this work.}
\affiliation{Department of Physics, Princeton University, Princeton, New Jersey 08544, USA}
\author{S.~L.~Sondhi}
\affiliation{Department of Physics, Princeton University, Princeton, New Jersey 08544, USA}

\begin{abstract}
Recent work has shown that a variety of novel phases of matter arise in 
periodically driven Floquet systems. Among these are many-body localized phases which 
spontaneously break global symmetries and exhibit novel multiplets of 
Floquet eigenstates separated by quantized quasienergies. Here we show 
that these properties are stable to {\it all} weak local deformations of 
the underlying Floquet drives---including those that explicitly break the defining symmetries---and that the models considered until now 
occupy sub-manifolds within these larger ``absolutely stable" phases. 
While these absolutely stable phases have no explicit global symmetries, they spontaneously break Hamiltonian 
dependent {\it emergent} symmetries, and thus continue to exhibit the novel multiplet structure. The multiplet structure in turn
encodes characteristic oscillations of the emergent order parameter at multiples of the fundamental period. Altogether these
phases exhibit a form of simultaneous long-range order in space and time which is new to quantum systems. We describe
how this spatiotemporal order can be detected in experiments involving quenches from a broad class of initial states.
\end{abstract}

\maketitle

\section{Introduction}\label{s:Intro}
The elucidation of phase structure is a major theme in condensed matter 
physics and statistical mechanics. An early paradigm for
doing so, associated most with Landau, characterizes phases through the spontaneous breaking of global symmetries present in the microscopic Hamiltonian {\it i.e}, phases are either paramagnetic, or spontaneously symmetry broken (SSB). 
In modern parlance, the phases obtained thereby are symmetry protected 
since their distinctions are erased if the symmetries are not present
microscopically. More recently, it has been found that this characterization is too coarse --- not all paramagnetic phases should be considered identical. Indeed, there exist 
paramagnetic symmetry protected topological (SPT) phases which
do not break any symmetries, but which nevertheless cannot be 
adiabatically connected to one another in the presence of the protecting 
global symmetry\cite{Chen10}. Remarkably, we now know of other phases, 
such as those with topological order, which do not even
require a global symmetry and are {\it absolutely stable}---their ground 
state (and sometimes even low temperature) properties are stable to arbitrary weak 
local perturbations\cite{WenNiu90,Kitaev03,Hastings05}. Equally remarkably, there 
are also examples of systems whose entire many body spectrum displays 
some absolutely stable property, namely many body localized \cite{Anderson58, Basko06, PalHuse, OganesyanHuse, Nandkishore14, AltmanVosk} (MBL) systems which 
robustly exhibit a full set of emergent local conserved 
quantities\cite{Huse14, Serbyn13cons, Imbrie2016, Serbyn13a, Chandran15b, Ros15,Pekker14,Rademaker2016}. One 
can also combine MBL with the above quantum orders to obtain MBL phases 
in which individual highly excited eigenstates show SSB, SPT, or 
topological order\cite{Huse13,PekkerHilbertGlass, Bauer13,Chandran14,Bahri15,Potter15}.

\begin{figure}
\includegraphics[width=\columnwidth]{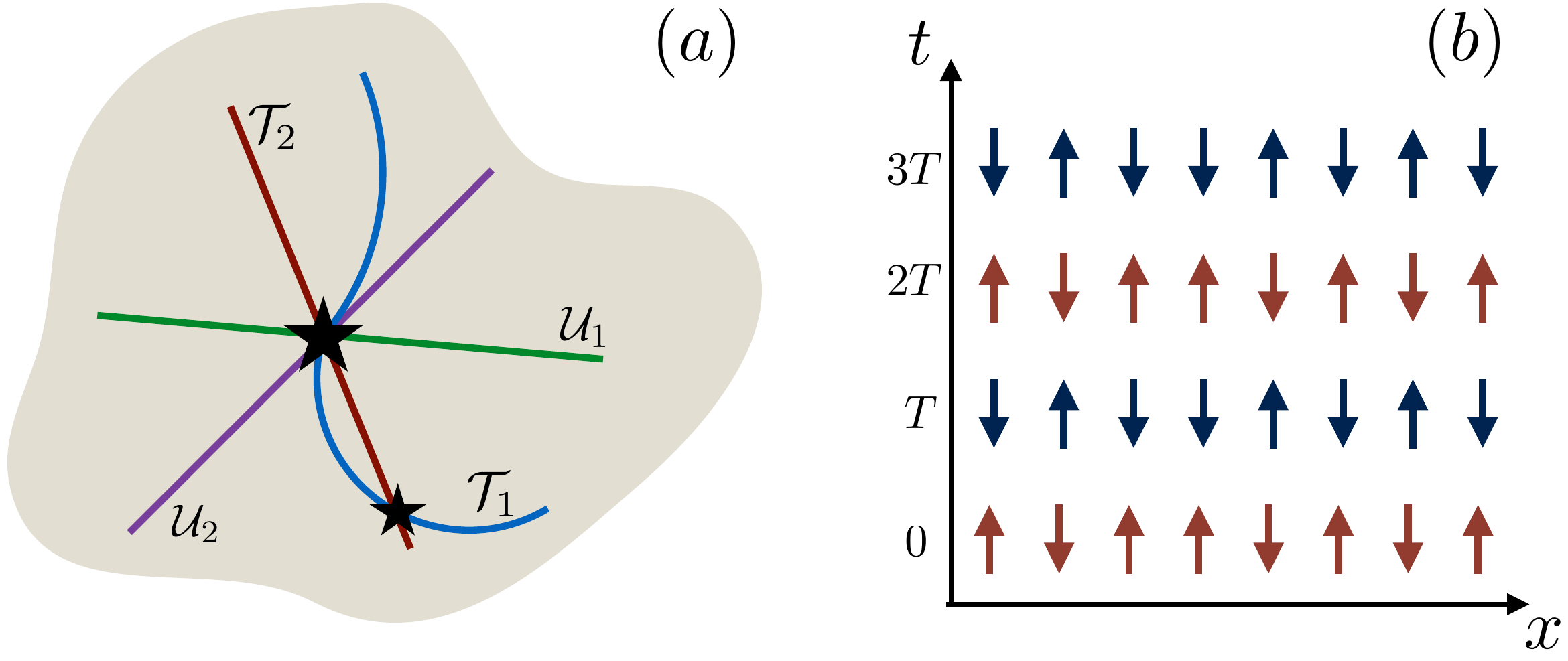}
\caption{(Color online): (a) Schematic depiction of the manifold of Floquet unitaries that are absolutely stable and characterized by Hamiltonian dependent emergent symmetries (grey area). Special sub-manifolds (colored lines) within the absolutely stable manifold are characterized by Hamiltonian {\it independent} unitary ($\mathcal{U}_i$) and antiunitary ($\mathcal{T}_i$) symmetries. Special models (black stars) can lie at the intersection of several sub-manifolds with exact symmetries. As an example, the $\pi$SG model defined in Refs.~\onlinecite{Khemani15, vonKeyserlingkSondhi16b} is absolutely stable and possesses the Ising unitary symmetry $P$ and an antiunitary symmetry $\mathcal{T} = KP$ where $K$ is complex conjugation. (b) Schematic depiction of the spatiotemporal long-range order found in absolutely stable phases---the order looks ``antiferromagnetic'' in time and glassy in space.}
\label{Manifolds}
\end{figure}

The ideas above assume  time translation invariance (TTI) or energy conservation since they involve describing the eigensystem of a time independent many body Hamiltonian. What happens if we relax this constraint, considering instead time dependent Hamiltonians $H(t)$? Generically, it is expected that an interacting, driven many-body system absorbs energy indefinitely and approaches a dynamic approximation to the infinite temperature equilibrium state. However,  for Floquet systems with periodic time dependence $H(t+T) = H(t)$, this fate can be avoided in the presence of sufficiently strong disorder (or in the absence of interactions\cite{Kitagawa10,Jiang11,Lindner11,Thakurathi13,Rudner13,Asboth14,Carpentier15,Nathan15,Roy15,Titum15a,Titum15b}) as was shown in recent work extending the physics of MBL to Floquet systems\cite{Lazarides14,Ponte15,Ponte15b,Abanin14,Rigol14}. This in turn allowed phases to be defined for MBL-Floquet systems\cite{Khemani15} via a generalization of the idea of eigenstate order first discussed for undriven MBL systems. In very recent work, a classification was given for phases that either preserve\cite{vonKeyserlingkSondhi16a,Else16,Potter16} or spontaneously break\cite{vonKeyserlingkSondhi16b} unitary global symmetries\footnote{We use the term ``spontaneously break unitary global symmetries''  to mean that the eigenstates exhibit the long-range order characteristic of spontaneous symmetry breaking.}.

In the present paper we build on the latter work and show that a subset of the SSB phases identified therein are stable to arbitrary weak local perturbations, including those that explicitly break any of the defining global symmetries. Thus this subset is {\it absolutely} stable---a remarkable outcome for a driven system.  The apparent
puzzle that SSB phases can be stable absent Hamiltonian independent symmetries is resolved elegantly: at general points in these absolutely stable phases, the drives (in the infinite volume limit) are characterized by a set of Hamiltonian {\it dependent} emergent unitary and antiunitary symmetries.
{\it Ex post facto}, we see that the symmetric models in Refs.~\onlinecite{Khemani15,vonKeyserlingkSondhi16b} live in lower dimensional submanifolds (characterized by Hamiltonian independent symmetries) of a much higher dimensional absolutely stable phase --- we sketch the resulting structure in Fig.~\ref{Manifolds}. This analysis uncovers a much richer symmetry structure than the global unitary symmetries used in previous work.

Strikingly, the out of equilibrium dynamics in these phases exhibits sharp universal signatures associated with oscillations of an emergent order parameter; these generalize the multiple period oscillations uncovered in previous work \cite{Khemani15,vonKeyserlingkSondhi16b} 
on symmetric drives. For example, we show that starting from arbitrary short range correlated initial states, the late time states show sharp oscillations of generic local operators at multiples of the fundamental period. This particular dynamical feature is a great boon to a future experimental detection of these phases as experimentalists are required neither to fine tune the Hamiltonian nor the starting state to observe a sharp signature! 

These longer periods raise the question of whether they should be thought of as representing spontaneous breaking of yet another symmetry---that of time translations by a period of the drive\footnote{We thank Ehud Altman for this incisive question.}. We note that
the idea that time translations might be analyzed in this fashion was first mooted by Wilczek \cite{Wilczek12} for time independent Hamiltonians; there is, however, now a proof \cite{Oshikawa15} that such ``time crystals'' do not exist for undriven systems in equilibrium.
We analyze this question further and find that strictly speaking {\it all} MBL systems, driven or undriven, exhibit some eigenstate correlations
characteristic of temporal {\it glasses}---an aperiodic breaking of time translation invariance (TTI). For the Floquet broken symmetry phases however,  the long distance correlations simultaneously exhibit spin glass order in space and multiple period oscillation in time. These lead
to the characteristic space-time snapshot illustrated for the simplest such phase in \figref{Manifolds}(b). Evidently the system exhibits spatiotemporal\footnote{While spatiotemporal order has been discussed for classical systems out of equilibrium, e.g. Ref.~\onlinecite{Sancho07}, to our knowledge this is the first appearance of such order for quantum systems.} long-range order in both space and time. The modulation in time, which is antiferromagnetic, does indeed break time translation symmetry but it preserves the combination of a translation and emergent Ising reversal. We note that a similar spatiotemporal order---now ferromagnetic in space---was previously exhibited in the large $N$ Floquet theory \cite{Chandran15} and discussed in the terminology of a lack of synchronization with the drive.

We note that the discovery of these absolutely stable Floquet phases can also be viewed as the realization that while a Hamiltonian that lacks \textit{any} symmetries (inclusive of time translation invariance) exhibits only a trivial phase, introducing discrete time translation invariance {\it alone} is sufficient to introduce a non-trivial phase structure. This would appear to be the minimum symmetry condition for this purpose.

In the rest of the paper we describe these results in more detail. We begin with the simplest example of a SSB phase that
is absolutely stable---this is the Ising $\pi$ spin-glass or $\pi$SG first described in Refs.~\onlinecite{Khemani15, vonKeyserlingkSondhi16b}.
In \secref{s:ases} we establish its absolute stability and analyze its emergent symmetries, correlations and characteristic spectral features within the
paradigm of eigenstate order. Next, in \secref{s:dynamics} we study the nature of dynamical correlations in the $\pi$SG 
in individual eigenstates and starting from generic short ranged entangled states, and discuss why the $\pi$SG should be identified as a Floquet
space-time crystal. We then discuss the catalog of other absolutely stable
Floquet phases in \secref{s:gen}, and show how some Floquet SPT phases exhibit time crystallinity at their boundaries.  We end with some concluding remarks in \secref{s:conclusion}. 

Before proceeding we note that a recent paper by Else, Bauer and Nayak \cite{Q} studies one of the submanifolds of our primary
example of an absolutely stable phase, the $\pi$SG and identifies it as a pure time crystal on the grounds that the drives break the unitary Ising symmetry. Our work clarifies that the order in the $\pi$SG and its cousins is always spatiotemporal and never purely temporal. Indeed the specific submanifold studied in \cite{Q} turns out to be protected by an antiunitary symmetry (see \eqnref{eq:antiunitaryelsenayaknew}) and thus exhibits spatial order in a particularly transparent form as we discuss below.

\section{ The $\pi$ Spin glass: Absolute stability and emergent symmetries}\label{s:ases}
We consider systems with time periodic local Hamiltonians $H(t)=H(t+T)$. The Floquet unitary is the time evolution operator for one period $U(T)\equiv\mathcal{T}e^{-i\int_{0}^{T}dtH(t)}$. The Floquet eigenstates $|\alpha\rangle$  of $U(T)$ have eigenvalues $e^{-i\epsilon_\alpha T}$, where $\epsilon_\alpha$ are the quasienergies defined modulo $2\pi/T$. Indeed, the Floquet eigensystem in phases with special forms of eigenstate order/quasienergy spectral pairing will form a central part of our discussion. 

\subsection{ Properties of the $\pi$SG phase}
Refs.~\onlinecite{Khemani15, vonKeyserlingkSondhi16a, vonKeyserlingkSondhi16b} discussed various SSB/SPT phases with Floquet eigenstate order, but not all of these phases are absolutely stable to arbitrary perturbations. In this work, our canonical example of an absolutely stable Floquet phase will be the $\pi$ spin-glass ($\pi$SG) phase\cite{Khemani15}. A concrete model Floquet unitary in this phase in 1d is    

\begin{equation}
\label{eq:piSG}
U_{f0}=P_x \exp[-i \sum_{r=1}^{L-1} J_r \sigma_r^z \sigma_{r+1}^z]; \,\,\,\,P_x = \prod_r \sigma_r^x,
\end{equation}
where $L$ is the system size, the $\sigma^\alpha_r$ for $\alpha = \{x, y,z\}$ are Pauli spin 1/2 degrees of freedom on site $r$,  $P\equiv P_x$ is the global Ising parity symmetry ($P_{y,z}$ analogously defined), and the $J_r$'s are random couplings drawn uniformly from $[\overline{J} - \delta J, \overline{J} + \delta J]$. We note several properties of this model, some of which were deduced in previous work\cite{Khemani15,vonKeyserlingkSondhi16b}:

\begin{enumerate}
\item $U_{f0}$ commutes with the unitary symmetry $P$. Defining anti-unitary operators $\mathcal{T}_\alpha = P_\alpha K$ where $K$ is complex conjugation, $U_{f0}$ also has $\mathcal{T} \equiv \mathcal{T}_x$ symmetry: $\mathcal{T} U_{f0} \mathcal{T}^{-1} = U^{-1}_{f0}$. It similarly has $\mathcal{T}_{y,z}$ symmetry for systems with an even number of sites\footnote{These symmetries have similar implications, which we do not discuss here for brevity.}. Thus, this model lies at the intersection of several special submanifolds with Hamiltonian independent symmetries (Fig.~\ref{Manifolds}) and is extremely robust to a large class of perturbations which preserve {\it some } exact symmetry. Note that the anti-unitary symmetries $\mathcal{T}$ are a combination of $K$ and a spatial Ising flip. 

\item The eigenspectrum of $U_{f0}$ can be found by noting that all the domain wall operators $D_r \equiv \sigma^z_r \sigma^z_{r+1}$ commute with $P_x$,  $U_{f0}$ and with one another. Thus, the eigenstates look like symmetric/antisymmetric global superposition states (also called cat states) of the form 
$$
|\pm\rangle \sim |\{d_r  \},p=\pm 1\rangle  = \frac{1}{\sqrt{2}}|\{\sigma_r^z\} \rangle \pm \frac{1}{\sqrt 2}|\{\overline{\sigma_r^z}\}\rangle\punc{,}
$$ 
where $\{\sigma_r^z\} = \{\uparrow \downarrow\downarrow \cdots \uparrow\}$ labels a frozen spin-glass configuration of $z$ spins (and hence the domain wall expectation values $d_r$), $\{\overline{\sigma_r^z}\}$ is its spin-flipped partner, and $p=\pm1$ is the Ising parity eigenvalue of the eigenstates.   

\item The eigenstates above have corresponding unitary eigenvalues $u(d, p) =  p e^{-i\sum_{r=1}^{L-1}J_{r}d_{r}}$.  Note that the opposite parity cat-state partners have unitary eigenvalues differing by a minus sign $u(d,-1)=-u(d,-1)$ and hence quasienergies differing by $\pi/T$. We refer to this phenomenon as a $\pi$ spectral pairing of cat states. 

\item The Floquet eigenstates exhibit long range connected correlations (LRO) and spin glass\cite{Fisher95,Vosk13} (SG) order in $\sigma^z_{i}$, but show no long-range order in $\sigma^x_{i}$ and $\sigma^y_{i}$. 

\item The order parameter for the $\pi$SG model oscillates with frequency $\pi/T$ or period $2T$, as indicated by the stroboscopic equation of motion $\sigma^z_r(nT)=(-1)^n \sigma^z_r$\cite{Khemani15,vonKeyserlingkSondhi16b}.  This follows directly from the fact that $\sigma^z_r$ anticommutes with $U_{f0}$. While $\langle \sigma^z_r (nT) \rangle =0$ in the Floquet eigenstates, the observable shows a periodic time dependence with period $2T$ in short-range correlated states of the form $ |\{\sigma_r^z\} \rangle \sim |+ \rangle + | - \rangle$. 
On the other hand, the $\sigma^x$ and $\sigma^y$ operators do not show period $2T$ oscillations.  
\end{enumerate}

\subsection{Absolute stability and emergent symmetries}

How robust are the above properties to perturbations of the form $H(t)\rightarrow H(t)+\lambda V(t)$? Numerical results  have already demonstrated the stability of $U_{f0}$ to weak Ising \cite{Khemani15} symmetric perturbations. 
We will provide evidence that this phase is, in fact, absolutely stable to {\it all} generic weak perturbations --- we will define dressed spin operators (Floquet l-bits) for the  perturbed system and show that it displays {\it emergent } symmetries with the same effect on eigenspectrum properties as the exact Ising symmetry.  

The first step in the argument is to observe that the stability of the localization of the unperturbed unitary to arbitrary weak local perturbations (for sufficiently strong disorder) is itself not a consequence of symmetries. More technically, call the corresponding perturbed Floquet unitary $U_{f\lambda}$ where $\lambda$ is the strength of the perturbation. We
expect that the stability of localization implies the existence of a family of local unitaries \footnote{A local (or low depth) unitary is a unitary which can be written as $\mathcal{V}=\mathcal{T} e^{-i \int^t_0 ds K(s)}$ for some local bounded Hamiltonian $K(t)$, with $t$ finite in the thermodynamic limit.} $\mathcal{V}_{\lambda}$ which relate the eigenvectors of $U_{f0}$ to those of $U_{f\lambda}$ for $\lambda$ in  some non-vanishing range \cite{Abanin14, Ponte15, vonKeyserlingkSondhi16b,Q}. Note that the locality of such a unitary is a subtle business outside of the very strongly localized region due to proliferating resonances and Griffiths effects\cite{Bauer13,Gopalakrishnan15}. 

Assuming that a low depth $\mathcal{V}_{\lambda}$ exists, it relates the new eigenvectors of $U_{f\lambda}$ denoted $|\alpha\rangle_\lambda $ to the eigenvectors of $U_{f0}$ via $$|\alpha \rangle_\lambda = \mathcal{V}_{\lambda} |\{d_r\},p\rangle.$$ The new quasienergies are similarly denoted as  $\epsilon^\alpha_\lambda$. These local unitaries allow us to define a set of dressed, exponentially localized operators $\tau_{r, \lambda}$ (analogous to the l-bits\cite{Serbyn13a,Serbyn13cons,Huse14,Ros15,Chandran15b} in static MBL systems) together with a dressed parity operator $P^{\lambda}$ via
\begin{align}\label{eq:l-bits}
\tau_{r,\lambda}^{\beta} &= \mathcal{V}_{\lambda} \sigma^\beta_{r}\mathcal{V}_{\lambda}^{\dagger} \nonumber \\
P^{\lambda} &= \prod_r \tau_r^x.
\end{align}
We will often suppress the explicit $\lambda$ dependence of $\tau^\alpha_{r,\lambda}$ for brevity and $\beta = {x,y,z}$. Defining (local) dressed domain wall operators as $D^\lambda_{r} \equiv  \tau_r^z \tau_{r+1}^z $, we get
\begin{align}
  D^\lambda_{r}|\alpha \rangle_\lambda &= \mathcal{V}_{\lambda}  (\sigma_r^z \sigma_{r+1}^z)  | \{d_r\},p\rangle = d_r |\alpha\rangle_\lambda, \nonumber \\
  P^{\lambda}|\alpha \rangle_\lambda &=  \mathcal{V}_{\lambda} P  | \{d_r\},p\rangle = p| \alpha \rangle_\lambda.
\end{align}
Thus, the perturbed eigenstates are also eigenstates of the dressed operators $ D^\lambda_{r}$ and $ P^{\lambda}$ which means these operators commute with $U_{f\lambda}$, and we can rewrite $|\alpha\rangle_\lambda$ more suggestively as  $|\{\tau_r^z\},p=\pm1\rangle$ using the same notation as before. By definition, $\tau_r^z$ anticommutes with $ P^{\lambda}$. Further we show in \appref{[Z:P]=pm1} that it also anticommutes with $U_{f\lambda}$ in the large system limit
\be\label{eq:tauAC}
 [ \tau_r^z, U_{f\lambda} ]_+ = O(e^{-c L}) \xrightarrow{L\rightarrow \infty} 0\punc{,}
\ee
using only the assumptions of locality and continuity. This implies that the Floquet eigenvalues are odd in $p$. Together with the previous statements about the commutation properties of $P^\lambda$ and $D^\lambda$ with $U_{f\lambda}$, it is easy to show that the unitary eigenvalues take the form ${u}_{\lambda}(\left\{  d_r \right\} ,p)=p e^{-if(\left\{  d\right\} )}$. Re-expressing the eigenvalue dependence on conserved quantities  in operator language gives
\be\label{eq:Ufcanonical}
U_{f\lambda}=P^{\lambda} e^{-if(\left\{  D^\lambda_{r} \right\} )}\punc{,}
\ee
where $f$ is a  functional of $D^\lambda$, or equivalently an even functional of the $\tau^z_r$'s. One can moreover argue that $f$ can be chosen to be local, using the fact that the Floquet unitary itself is low depth\cite{vonKeyserlingkSondhi16a,vonKeyserlingkSondhi16b}. Thus, $f$ generically takes the form
$$
f(\{D^\lambda\}) = \sum_{ij} J_{ij}\tau_i^z \tau_j^z + \sum_{ijkl} J_{ijkl} \tau_i^z \tau_j^z \tau_k^z \tau_l^z + \cdots
$$
where the couplings $J_{ij} \sim e^{-|i-j|/\xi}$ decay exponentially with distance reflecting the locality of the unitary. 

Written this way, the Floquet unitary \eqref{eq:Ufcanonical} clearly has a $\mathbb{Z}_{2}$ symmetry $P^{\lambda}$ --- although we say it is emergent because $ P^{\lambda}$, in general, depends on the details of the underlying Hamiltonian. $U_{f\lambda}$ similarly has an emergent antiunitary symmetry $\mathcal{T}^\lambda \equiv  P^{\lambda} K^\lambda$ where $K^\lambda$ is complex conjugation defined with respect to the $\tau^\alpha$. Note that \eqnref{eq:Ufcanonical} takes much the same functional form as the model unitary \eqnref{eq:piSG}, and correspondingly its eigenstates exhibit long-range order in the dressed order parameter $\tau^z_r$ (associated with spontaneous breaking of $P_\lambda$), and short range order in $\tau^{x,y}_r$. The statements about $\pi$ spectral pairing and the temporal dependence of observables (in particular  $\tau^z(nT)=(-1)^n \tau^z(0)$) also follow directly\footnote{In principle we can now identify symmetry defined submanifolds based on keeping the emergent symmetries about {\it any} fixed point in the $\pi$SG which provides an ``origin independent'' view of the structure of the phase. The functional form of the perturbed unitary $U_{f\lambda}$ \eqref{eq:Ufcanonical} and its implications are among the central results of this paper.}. 

Finally, we note that Refs~\onlinecite{Khemani15,vonKeyserlingkSondhi16a} also defined a $0$SG phase with the model unitary $\exp[-i \sum_{i=1}^{L-1} J_i \sigma_i^z \sigma_{i+1}^z]$. Like the $\pi$SG, this is also a phase with long-range SSB Ising order, but one in which the cat states are degenerate instead of being separated by $\pi/T$. If we generically perturb about this drive, we must begin with Floquet eigenstates that explicitly break the Ising symmetry in order for the change of basis unitary $\mathcal{V}_\lambda$ to be local. Implicitly this requires us to work in the infinite volume limit directly. In this case, one can show that $\tau^z$ commutes (rather than anticommutes) with the Floquet unitary, and one can readily use this to split the degeneracy between the Floquet eigenstates, rendering this phase unstable to arbitrary perturbations.  By contrast, in the $\pi$SG phase, the cat states are $\pi$ split and therefore non-degenerate --- a fact which is essential to the stability of the SSB order to arbitrary perturbations.  

\subsection{Long range order and numerics}

\begin{figure}
\includegraphics[width=\columnwidth]{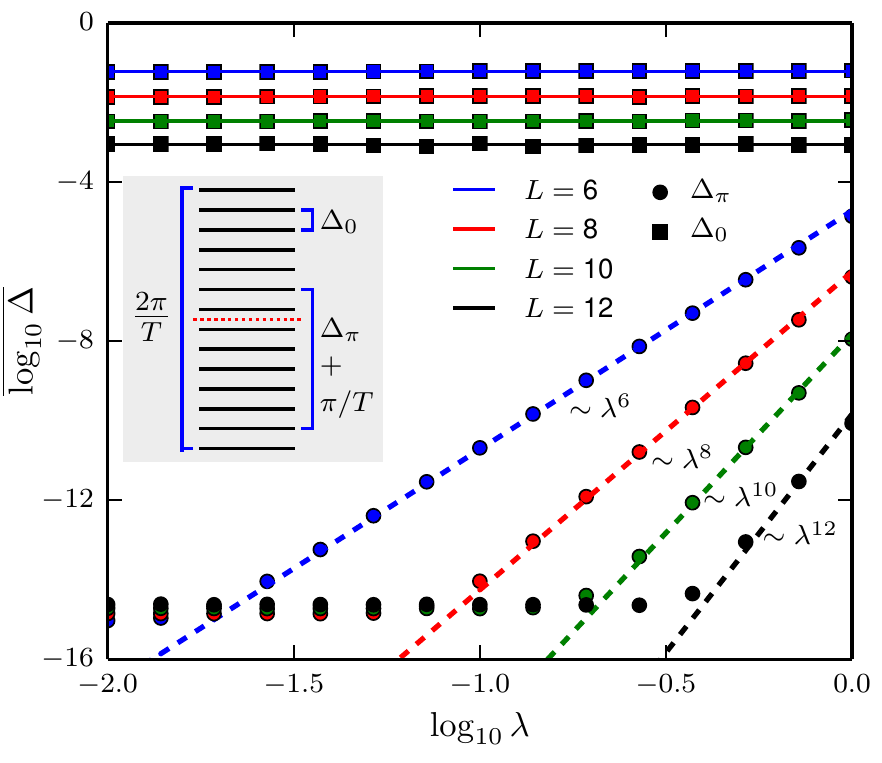}
\caption{(Color online): Disorder and eigenstate averaged spectral gaps for the generically perturbed model \eqref{eq:perturbedU} without any $P$ and $\mathcal{T}$ symmetries plotted as a function of the perturbation strength $\lambda$ and system size $L$. The nearest-neighbor quasienergy gap $\Delta_0$ shows no $\lambda$  dependence but decreases exponentially with $L$. On the other hand $\Delta_\pi$ which measures the spectral pairing of even-odd parity states scales as $\lambda^L$ (fits to this form superimposed). Thus, there is a window of $\lambda$s for which $\Delta_\pi \ll \Delta_0$ and the system exhibits robust spectral pairing in the $L\rightarrow \infty$ limit. 
Gaps smaller than $\sim 10^{-14}$ are below numerical precision, thus the initial $\lambda$ independent trend in the $\Delta_\pi$ data for larger $L$. (inset): Cartoon of the quasienergy spectrum illustrating the definitions of $\Delta_0$ and $\Delta_\pi$.  }
\label{Pairing}
\end{figure}

We now numerically check for the predicted $\pi$ spectral pairing in a perturbed model of the form 
\begin{equation}
\label{eq:perturbedU}
U_{f\lambda} = P \exp[ -i \sum_{r=1}^{L-1} J_r \sigma_r^z \sigma_{r+1}^z  -i\lambda \sum_{r=1}^L h_r^x \sigma_r^x + h_r^y \sigma_r^y +h_r^z \sigma_r^z]
\end{equation}
The fields $J_r, h_r^{x,y,z}$ are drawn randomly and uniformly with $\overline{J_r} = 1, \delta J_r = 0.5, \; \overline{h_r^x}= \delta h_r^x = 0.1, \; \overline{h_r^y}= \delta h_r^y = 0.15, \; \overline{h_r^z}= \delta h_r^z = 0.45$ and the notation $\overline{x}, \delta x$ means that $x$ is drawn from $[\overline{x}-\delta x, \overline{x} + \delta x]$. The perturbation breaks all the unitary and anti-unitary symmetries present in the original $U_{f0}$ model. To check for spectral pairing, we define the nearest neighbor gap between the perturbed quasienergies as $\Delta_0^i = \epsilon^\lambda_{i+1} - \epsilon^\lambda_i$ and the $\pi$ gap as $\Delta_\pi^i = \epsilon^\lambda_{i+ \mathcal{N}/2} - \epsilon^\lambda_i - \pi/T$ where $\mathcal{N}=2^L$ is the Hilbert space dimension, and where the second equation follows from the fact that the  quasienergy bandwidth is $\pi/T$ and we expect states halfway across the spectrum to be paired at $\pi/T$ (See Fig.~\ref{Pairing} (inset) for an illustration of these definitions). The system shows spectral pairing at $\pi$ if there is a range of $\lambda$'s for which $\Delta_\pi \ll \Delta_0$ as $L \rightarrow \infty$. Fig.~\ref{Pairing} shows the mean $\Delta_\pi$ and $\Delta_0$ log-averaged over eigenstates and several disorder realizations for different $\lambda$'s and $L$'s. We see that $\overline{\Delta_\pi} \sim \lambda^L$ whereas $\overline{\Delta_0} \sim e^{-sL}$ where $s \sim \log(2)$ is a $\lambda$ independent entropy density. Thus, we can get robust pairing in the window $|\log{\lambda}| > s$. 

\begin{figure}
\includegraphics[width=\columnwidth]{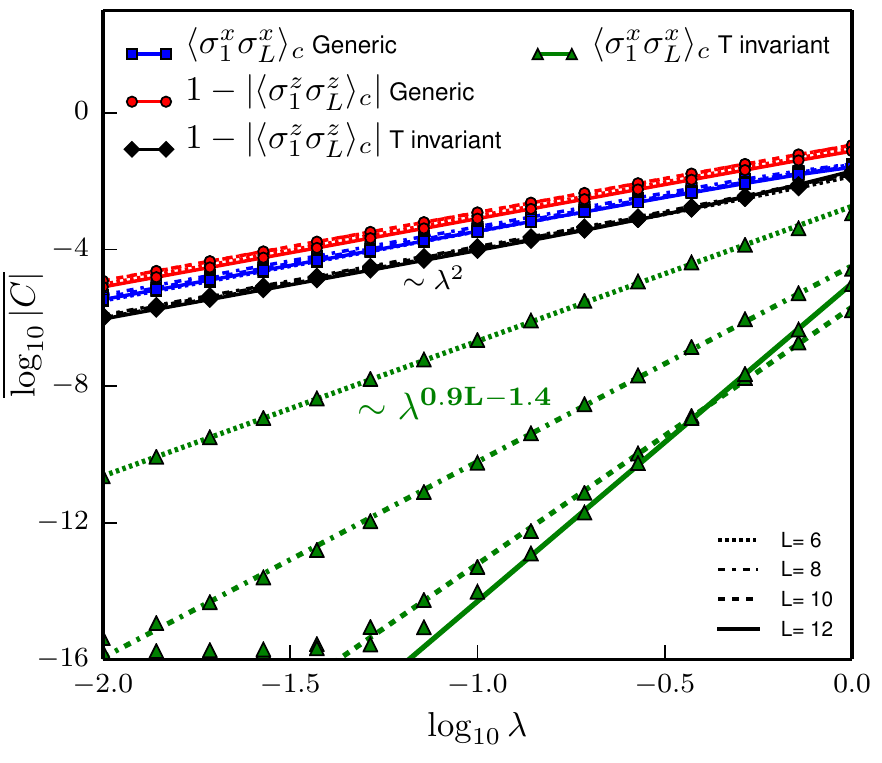}
\caption{(Color online): Disorder and eigenstate averaged end-to-end connected correlation functions for $\sigma^{x,z}$ in the ``generic'' model \eqref{eq:perturbedU} with no $P,\mathcal{T}$ symmetries (blue squares, red circles) and a model\cite{Q} with $\mathcal{T}$ symmetry obtained by setting $h^y =0$ in \eqref{eq:perturbedU} (black diamonds, green triangles). As discussed in the text, the generic model shows long-range order for both operators which is signaled here by correlations scaling as $\lambda^2 $ independent of system size. On the other hand, in the model with $\mathcal{T}$ symmetry, only $\sigma^z$ shows long-range order while the $\sigma^x$ correlator scales as $\lambda^{f(L)}$  where $f(L) \sim 0.9L - 1.4$ (fits shown) and thus vanishes in the $L\rightarrow \infty $ limit. This is to be expected from symmetry constraints. The $\sigma^y$ correlators (not shown here) also display long-range order in the generic model but not in the $\mathcal{T}$ symmetric model. }
\label{corr}
\end{figure}

Having shown how the robustness of the $\pi$SG phase is associated with spontaneously broken emergent symmetries and long-range order in the $\tau^z$ variables, we can now ask what effect this long-range order has on correlations in the physical $\sigma^\alpha$ degrees of freedom. Generically we expect the expansion of the physical spins in terms of l-bits to have some components which are diagonal and odd in $\tau^z$, for example  $\sigma^\alpha_r=c^\alpha \tau^z_{r}+\cdots$. As a result $\sigma^{\alpha=x,y,z}_r$ are all expected to have long range connected correlation functions, as well as a component exhibiting $2T$ periodic stroboscopic oscillations. These predictions agree with our numerical results \figref{corr} and \figref{Rabi} respectively.

On the other hand, when we perturb $U_{f0}$ in a manner that respects an explicit symmetry like $P$ or $\mathcal{T}$, the resulting models reside in a special submanifold of the absolutely stable phase. The presence of the exact symmetries constrains the form of the dressed $\tau^\alpha$ operators and leads to concrete predictions about the order in and temporal dependence of different operators.  For example, it was argued\cite{vonKeyserlingkSondhi16b} that when the perturbation $\lambda V(t)$ is such that $U_{f\lambda}$ continues to have Ising symmetry, $\mathcal{V}_\lambda$ can be chosen to commute with $P$. As a result, $ P^{\lambda} = P$, and $\sigma^{y,z}$ are odd under $ P^{\lambda}$  whereas $ \sigma^x$ is even under $  P^{\lambda}$. This means an operator expansion of $\sigma^x_r$ in terms of the dressed $\tau^\alpha_r$ operators can only involve even combinations of $\tau$:  $\sigma_r^z = \alpha_1 \tau_r^x+\beta_2 \tau^z_{r} \tau^z_{r+1} +\ldots $. Hence the {\it connected} correlation functions of $\sigma^x_{rs}$ should decay exponentially with $|r-s|$, and this operator is not expected to have robust period $2T$ oscillations. On the other hand $\sigma_r^{y,z}$ will generically exhibit both long range connected correlations as well as period $2T$ oscillations. 

Similarly we can pick perturbations for which $U_{f,\lambda}$ respects antiunitary symmetries like $\mathcal{T}= P K$, i.e., for which $\mathcal{T} U_{f,\lambda} \mathcal{T} =   U_{f,\lambda}^{-1}$. As an example, the model studied in Ref.~\onlinecite{Q} resembles \eqnref{eq:perturbedU}  with $h^y=0$, so has the effect of perturbing \eqnref{eq:piSG} by  $\lambda V  \sim h_r^z \sigma_r^z + h_r^x \sigma_r^x$. With this choice of $V$ it is straightforward to verify that the corresponding $U_{f,\lambda}$ respects $\mathcal{T}$ symmetry
\begin{align}\label{eq:antiunitaryelsenayaknew}
&\mathcal{T} U_{f,\lambda} \mathcal{T}^{-1} =   (P K) U_{f\lambda} (P K)^{\dagger}  \nonumber \\
&= (P K) P \exp[-i \sum_{r=1}^{L-1} J_r \sigma_r^z \sigma_{r+1}^z  +h_r^z \sigma_r^z + h_r^x \sigma_r^x]  (P K)^{\dagger} \nonumber \\ 
 &=  \exp[i \sum_{r=1}^{L-1} J_r \sigma_r^z \sigma_{r+1}^z +h_r^z \sigma_r^z + h_r^x \sigma_r^x ] P \nonumber \\
 &=U^{-1}_{f,\lambda}\punc{.}
\end{align}
 In this case, we can pick the change of basis matrix $\mathcal{V}_\lambda$ to commute with $\mathcal{T}$ (see \appref{app:symmetries Vl}) which implies that $\tau^x,\tau^y, \tau^z$ are even, even, and odd respectively under $\mathcal{T}$. In turn, the operator expansions of $\sigma^{x,y}$ can only contain terms with even numbers of $\tau^z$s in their expansions. Hence neither should exhibit protected $\pi/T$ oscillations, nor should they have long range connected correlations as demonstrated in \figref{corr}. This accounts for the absence of $\pi/T$  oscillations for $\sigma^x_r(nT), \sigma^y_r(nT)$ in the data presented in Ref.~\onlinecite{Q}.

\section{The $\pi$ Spin Glass: Spatiotemporal Long Range Order}\label{s:dynamics}
We have already discussed above that at general points in the absolutely stable $\pi$SG phase the emergent order parameter operators,
$\tau^z_i$, change sign every period. Prima facie, this implies the spatiotemporal order sketched in Fig.~1b: spin glass order in space and antiferromagnetic order in time.

 The aim of this section is to more sharply characterize this spatiotemporal order. As the $\pi$SG is a localized phase, unlike in the equilibrium context, there is not an obviously correct set of correlations one should examine to detect said order. We propose to examine the time dependent one and two point correlation functions of local operators in two families of states. The first are the Floquet eigenstates which are the basis of the eigenstate order paradigm of phase structure in Floquet systems. The second are the late time states reached by time evolving from general initial states; these are particularly relevant to experiments where the preparation of Floquet eigenstates is not feasible.
\subsection{Eigenstate correlations and response}

We start by considering Floquet eigenstates for the $\pi$SG.  All single time operators $\langle O(t) \rangle$ in these are strictly 
periodic with period $T$---this is the analog of the time independence of single time operators in Hamiltonian eigenstates and hence the temporal component of the order is invisible to such operators. The invisibility of temporal order in the $\langle O(t) \rangle$ is analogous to the invisibility of Ising symmetry breaking in one point expectations of spatially local Ising-odd operators in globally Ising symmetric states. From this perspective\cite{Oshikawa15} it follows that to detect temporal order we must either  (a) examine a two time function of some operator or (b) explicitly add an infinitesimal field that selects the desired temporal order (much as we would examine long-range order in two-point functions of Ising-odd variables and/or add an infinitesimal Ising symmetry breaking term to detect spontaneously broken Ising symmetry).

We begin with (a) and examine time-dependent correlators 
\begin{align}
C_\alpha (nT; r,s) &\equiv \langle \alpha| O_r(nT) O_s |\alpha \rangle \nonumber \\
& = \sum_\beta  e^{-in T(\epsilon_{\alpha} - \epsilon_{\beta})}\langle \alpha| O_r|\beta \rangle \langle \beta| O_s |\alpha \rangle
\end{align}
of operators $O_{r/s}$ localized near sites $r,s$ in the Floquet eigenstates $|\alpha \rangle = |\{d\}, \pm\rangle_\lambda$ (see \secref{s:ases} for notation). The operator expansion of $O_{r/s}$ in the $\tau^\alpha$ basis will generically contain terms that are odd combinations of $\tau^z$s. In the $\pi$SG phase, these have matrix elements between $|\alpha\rangle$ and its parity flipped partner and thus $ C_\alpha (nT)$ generically has a frequency $\pi/T$ component. In addition, the off-diagonal terms in the operator expansion involving $ \tau^{\{x,y\}} $ will make local domain wall excitations near sites $r/s$. Now a crucial point: if $r,s$ are held a fixed distance apart in the infinite volume limit, then $C_\alpha (nT)$ breaks TTI for {\it any} MBL-Floquet system. The reason is that one can crudely view a Floquet MBL system as a set of weakly interacting localized modes (the effective domain wall operators in this case) each with their own local spectra.  As in the simplest case of 2-level systems whose physics is that of Rabi oscillations, these local subsystems (which are excited by $\tau^{x/y}$) exhibit response at frequencies incommensurate with the driving frequency. The presence of these incommensurate frequencies means $C_\alpha(nT)$ in {\it all MBL-Floquet systems always look glassy}, although for the $\pi$SG there is generically also a quantized response at $\pi/T$. 

This short distance temporal glassiness however goes away when we examine long distances in space by placing the operators arbitrarily far apart in an infinite system, {\it i.e.}, by taking $\lim_{L \rightarrow \infty}$ before examining the limit $|r-s| \rightarrow\infty$.  Since the operator expansions of $O_{r/s}$ are exponentially localized near sites $r/s$,  the off-diagonal terms in the expansion of $O_r$ which create domain-wall excitations near site $r$ {\it cannot be annihilated} by the action of $O_s$ in the limit $|r-s| \rightarrow\infty$ under the assumption of locality.

Thus, the only terms that contribute to 
$C_\alpha(nT; r,s)$ in this limit are diagonal in $\tau^{z}$s. Terms odd in 
$\tau^{z}$ give a response at $\pi/T$ while the even terms give a 
response at frequency $0$. Thus we can write
$$C_\alpha(nT; r,s) \sim c_0(r;\alpha) c_0(s;\alpha)  + c_1(r;\alpha) c_1(s;\alpha) (-1)^n$$
where the second piece reflects the spatiotemporal order of the odd $\tau^z$ terms, as well as  the connected part of the correlation function. The dependence of the coefficients on $r$, $s$ and $\alpha$ has been made explicit to emphasize the glassy nature of the order in space. This establishes a connection between the long range spatial order in the eigenstates and the period $2T$ temporal order.

The above analysis can be complemented by taking the approach (b) and adding to $H(t)$ a ``staggered field'' in time of the form
$\epsilon \sum_n (-1)^n V \delta(t -nT)$, where $V$ is odd and diagonal in $\tau^z$.
Now consider time-dependent expectation values of generic local operators $O_r$ (which have a projection on odd $\tau^z$ terms) in the Floquet eigenstates $|\alpha\rangle_\epsilon$ for the new period $2T$ unitary which can be reshuffled to the form
$U_{f,\epsilon}(2T)=e^{- i 2\epsilon V} U^2_{f,0}$. This problem looks like the classic Ising symmetry breaking problem. At $\epsilon=0$,
$U_{f,\epsilon}(2T)=U^2_{f,0}$ has two degenerate states in the infinite volume limit. If $V$ breaks the symmetry between two members of the  doublet then
 $$  \lim_{ \epsilon \rightarrow 0} \lim_{ L \rightarrow \infty} {}_\epsilon\langle \alpha |O_r(nT)| \alpha \rangle_\epsilon = b_0(r;\alpha) + b_1(r;\alpha) (-1)^n $$
since the perturbed period $2T$ eigenstates $|\alpha\rangle_\epsilon$ just look like product states of $\tau^z$ in this limit and are thus superpositions of the opposite parity eigenstates of $U_{f\lambda}$.  On the other hand, the opposite order of limits gives $\lim_{ L \rightarrow \infty}   \lim_{ \epsilon \rightarrow 0} {}_\epsilon\langle \alpha |O_r(nT)| \alpha \rangle_\epsilon =  b_0(r;\alpha)$.  
We emphasize that the measures discussed here are {\it eigenstate} measures. If averaged over all eigenstates the signatures vanish.

\subsection{Quenches from general initial states}\label{ss:dynamicsSRE}

We now turn to the question of evolution from more general initial states rather than eigenstates. This is experimentally important, and more particularly so because the Floquet eigenstates for the $\pi$SG are macroscopic superpositions and thus hard to prepare. For concreteness, consider starting from a short-range correlated state like a product state of the physical spins. In the following we will adapt the analysis
of dephasing in quenches in MBL systems \cite{Serbyn13a,Huse13}. We will assume that the starting state exhibits
a non-zero expectation value for the order parameter, i.e. $\langle \psi_0|\tau^z_i|\psi_0 \rangle \ne 0$; if it does not the temporal
features will be entirely absent.  For simplicity we will only discuss one point functions as they are already non-trivial in this setting and the generalization is straightforward.

\begin{figure}
\includegraphics[width=\columnwidth]{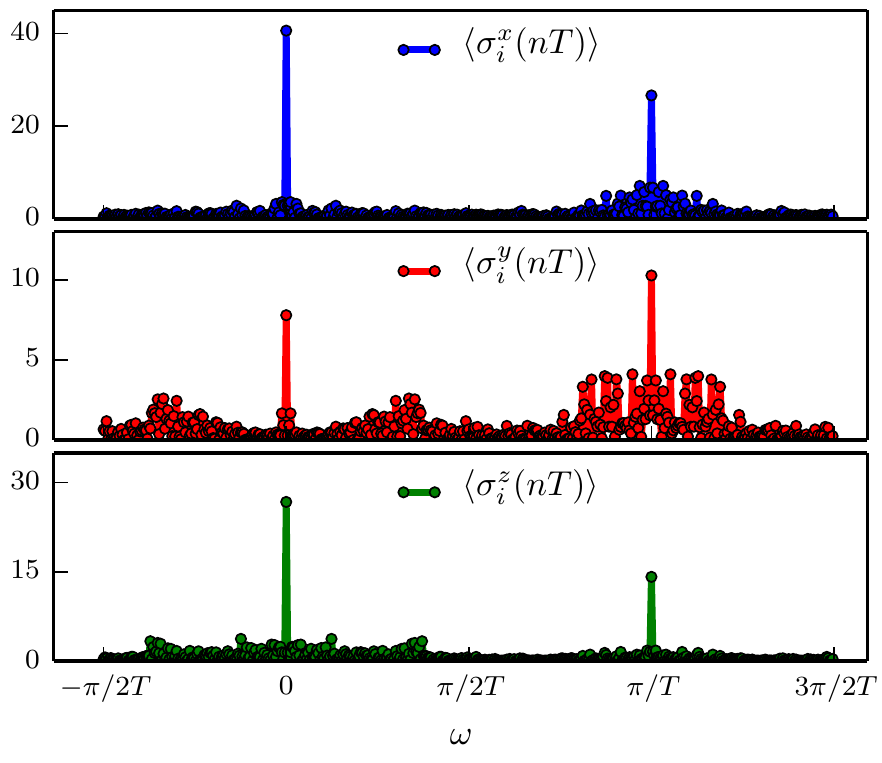}
\caption{(Color online): Fourier transform over time window $\Delta t=500 T$ of one point time-dependent expectation values $\langle \psi_0|\sigma^{\{x,y,z\}}(nT)|\psi_0\rangle$ in the ``generically'' perturbed model \eqref{eq:perturbedU}. The initial state $|\psi_0\rangle$ is a product state with physical spins $\sigma^\alpha$ randomly pointing on the Bloch sphere and uncorrelated from site to site. As discussed in the text, the response looks ``glassy'' with several incommensurate Fourier peaks in the addition to the peak at $\pi/T$, although we expect these to decay away in the $L\rightarrow \infty,  T \rightarrow \infty$ limit. Data is shown for a single disorder realization in a system of length $L=10$.   }
\label{Rabi}
\end{figure}

In a finite size system, $\tau^z$ only anticommutes with the Floquet unitary up to exponentially small in $L$  corrections \eqref{eq:tauAC}, which in turn introduce corrections to the equation of motion: $ \tau^z(nT) = (-1)^n\tau^z(0) + O(e^{-L})$. This leads to exponentially small shifts in the spectral pairing at $\pi/T$ which varies randomly between pairs of eigenstates. Ignoring these shifts for
times $1 \ll t \ll O(e^{+L})$, one can readily show that for any finite system the one point functions will generically show glassy behavior with incommensurate Fourier peaks along with an additional peak at $\pi/T$; see \figref{Rabi} for an illustration. More precisely, the logarithmic in time dephasing of correlations in MBL systems\cite{Serbyn13a,Huse14} can be used to show that the correlators will show aperiodic behavior stemming from these additional Fourier peaks with a power law envelope $\sim t^{-b}$, where $b>0$ depends on the localization length\cite{Serbyn13a}. {\it Thus, finite systems at large but not exponentially large times look like time-glasses with an additional quantized response at $\omega = \pi/T$. }  However, if one waits a time $t \sim e^{L}$ that is long enough to (i) resolve the exponentially small many-body level spacings and (ii) to resolve the shifts in the spectral pairing away from $\pi/T,$ both the peak at $\pi/T$ and the extra incommensurate peaks almost entirely decay away due to usual dephasing mechanisms leaving behind aperiodic oscillations with a magnitude of $O(e^{-L})$. It is worth reminding the reader that the precise details of the time dependence will reflect the choice of the starting state and disorder realization.

We can formalize the above in two non-commuting limits: (a) $\lim_{ t\rightarrow \infty} \lim_{ L\rightarrow \infty}$ and (b) $\lim_{L\rightarrow \infty} \lim_{ t\rightarrow \infty}$. While (a) characterizes the ``intrinsic'' quench dynamics of this phase, experiments will only have access to limit (b). In (b) the late time aperiodic oscillations with envelope $O(e^{-L})$ discussed above also go away, and the one-point functions are constants. In (a) we never reach times of $O(e^{L})$ and instead observe persistent oscillations with period $2T$ out to $t \rightarrow \infty$ with all additional incommensurate oscillations decaying away as a power of time. 

Thus, the intrinsic dynamical response of this phase is characterized by a single quantized Fourier peak at $\omega = \pi/T$ which goes along with formally exact spectral pairing at $\pi/T$ and LRO in $\tau^z$. In this limit, the late time state exhibits a precisely doubled period for every single realization of disorder and combined space-time measurements would lead precisely to the kind of snapshot sketched in Fig.~1b. More concretely, state-of-the-art experiments in ultracold atoms \cite{Schreiber2015, Kondov2015,Bordia2016, Choi2016}  have convincingly demonstrated that a fingerprint of the initial state persists to asymptotically late times in the MBL phase. In a  generalized experimental setup probing the $\pi$SG phase in the MBL Floquet problem \footnote{We thank Christian Gross for discussions on possible experiments.},  the persistence of the starting fingerprint would measure localization and spatial spin glass order, while oscillations in time would measure the temporal response at $\pi/T$. We also note that a recent experiment demonstrated signatures of MBL in two dimensions\cite{Choi2016} and, more generally, we expect our considerations to apply in all dimensions where MBL exists\cite{Chandran20162D}. 

\subsection{Comments}

In the above discussion we have considered two settings, that of Floquet eigenstates and of late time states stemming from quenches. It is useful to contrast our findings with their analogs for general MBL phases (Floquet or undriven), and for ETH obeying phases (focussing on the undriven case, as the
Floquet version has trivial infinite temperature correlations). We find that unequal time correlations in eigenstates generically break TTI in all MBL phases, which thus generically look glassy. By contrast similar correlations in ETH systems do {\it not} generically break TTI. In the $\pi$SG we find that eigenstate correlations specifically designed
to pick out the order parameter dynamics are ``antiferromagnetic'' in the time domain and thus break TTI while they are ``ferromagnetic'' for the $0$SG and thus do not. Turning now to the late
time states coming from quenches, in MBL phases these are initial state dependent while in ETH phases these are not. Hence if we look for TTI breaking via these late time states we do not observer it in all ETH phases as well as MBL phases except the $\pi$SG (and its relatives which we discuss in the next section). We remind the reader though that in the $\pi$SG we need to quench from states that exhibit a macroscopic expectation value for the order parameter. All in all we conclude that the $\pi$SG exhibits a distinct and novel pattern of spatiotemporal order that is new to quantum systems.

\section{Generalizations}\label{s:gen}
Here we list a number of generalizations of the $\pi$SG phase. Ref.~\onlinecite{vonKeyserlingkSondhi16b}  presented a family of models with an explicit global symmetry group $G$ which exhibit eigenstate long-range order,  protected spectral pairing and temporal crystallinity.  First we note that, much like the $\pi$SG, many of these models are absolutely stable to local perturbations, even those that break the global symmetry $G$. We then explain why bosonic SPT Floquet drives\cite{vonKeyserlingkSondhi16a,Else16,Potter16,Harper16} are not stable to the inclusion of symmetry breaking perturbations, although in the presence of the protecting symmetry they exhibit time crystallinity at their edges.

\subsection{ $\mathbb{Z}_{n}$ and non-abelian models: }
Consider first models with global $\mathbb{Z}_n$ symmetry\cite{vonKeyserlingkSondhi16b,Rehn16}. There are $n$ possible phases with completely spontaneously broken symmetry\cite{vonKeyserlingkSondhi16b}, labelled by $k=0,1,\ldots n-1$. The eigenvectors of the corresponding unitary are the $\mathbb{Z}_{n}$ equivalents
of cat states i.e., macroscopic superpositions of $n$ spin configurations. In cases with $k\neq0$, and in the presence of $\mathbb{Z}_{n}$ symmetry, the spectrum consists of multiplets of $n$ cat states appearing in $n/g$ distinct groups each with degeneracy $g\equiv\text{gcd}\left(n,k\right)$. The $n/g$ distinct groups are split by quasienergy multiples of $2\pi g/nT$. As for the $\pi$SG, some
of these statements survive even when $\mathbb{Z}_{n}$
symmetry is explicitly broken. In particular, while the $g$ fold degeneracy for each group of cat states can readily be broken,  it remains the case that each eigenstate is paired in a multiplet of $n/g$ related cat states, separated by quasienergy $2\pi g/nT$. A similar statement holds for the non-abelian models in Ref.~\onlinecite{vonKeyserlingkSondhi16b}. These more general drives have an explicit unitary non-abelian symmetry $G$,
and are classified by an element of the center of the group $z\in Z(G)$. Let $q$ denote the order
of $z$. The spectrum consists of $q$ groups of $G/q$ degenerate cat-like states, and  the $q$ groups are separated by quasi-energies which are multiples of  $2\pi/qT$. The $|G|/q$ degeneracy at each quasienergy can once again be lifted using symmetry breaking perturbations, but each eigenstate is still paired with $q$ cat state partners, split by quasienergy multiples of $2 \pi / qT$.

\begin{figure}
\includegraphics[width=\columnwidth]{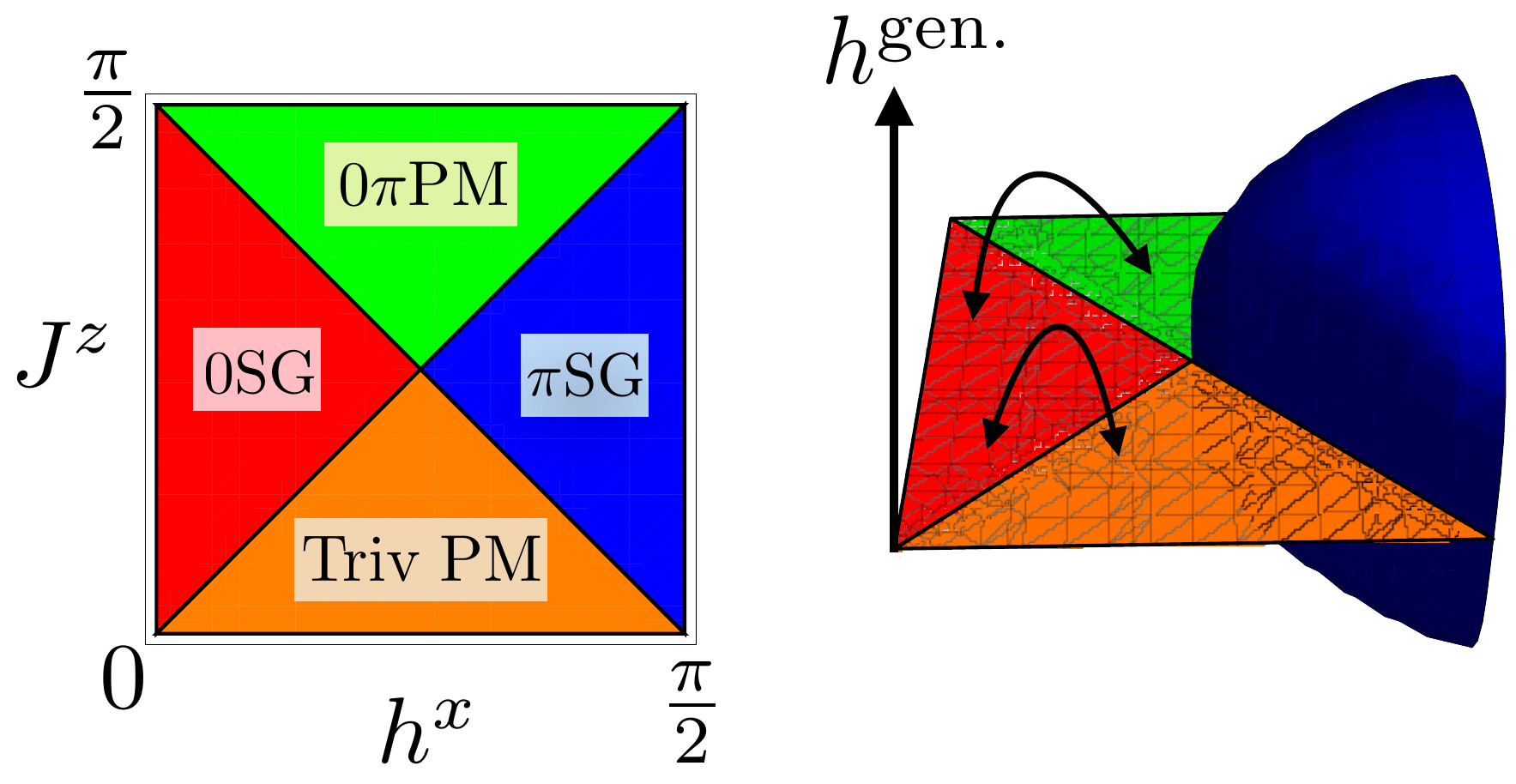}
\caption{(Color online): (left) Phase diagram for the MBL Ising symmetric drives presented in Refs. \onlinecite{Khemani15, vonKeyserlingkSondhi16b} showing the $0$SG and $\pi$SG phases which are long-range ordered and spontaneously break Ising symmetry, as well as the $0\pi$PM and trivial paramagnetic phases which have no LRO. The $0\pi$PM is an SPT with non-trivial edge modes and can spontaneously break time translation symmetry on its edges. (right): On perturbing with generic Ising symmetry breaking fields $h^{\rm gen}$, only the $\pi$SG is absolutely stable and continues into a phase with LRO and an emergent symmetry. The other three phases can be continuously connected to the trivial MBL paramagnet in the presence of $h^{\rm gen}$.  } 
\label{Perturbed}
\end{figure}

\subsection{ Stability of SPTs and boundary time crystallinity} While the $\pi$SG phase
is absolutely stable, similar Floquet generalizations of bosonic SPT
phases\cite{vonKeyserlingkSondhi16b} are not. Before showing this, let us first note that some Floquet SPTs spontaneously break TTI at their boundaries. This boundary TTI breaking is not tied to bulk LRO and the phases are correspondingly {\it unstable} to symmetry breaking perturbations. 
We illustrate this with the simple example of an Ising Floquet SPT, the so-called $0\pi$PM\cite{Khemani15, vonKeyserlingkSondhi16a}. In fact, the $0\pi$PM and $\pi$SG are neighbors on a common Floquet phase diagram\cite{Khemani15, vonKeyserlingkSondhi16a,vonKeyserlingkSondhi16b} \figref{Perturbed}(left) which also contains the $0$SG discussed earlier and a trivial MBL paramagnet. A simple Floquet unitary for $0\pi$PM on a system with boundary is \cite{ vonKeyserlingkSondhi16a} 
\begin{equation}\label{eq:0 pi PMbdry}
U_{f} =\sigma^z_1 \sigma^z_N \exp[-i \sum^{N-1}_{r=2} h_r \sigma^x_r],
\end{equation}
where the fields $h_r$ are randomly distributed. This model has trivial bulk paramagnetic eigenstate order, but it also has non-trivial Ising odd  ``pumped charges'' $\sigma^z_{1/L}$, using the parlance of Ref.~\onlinecite{vonKeyserlingkSondhi16a}. As a consequence, the eigenspectrum exhibits ``spectral quadrupling''. Labeling the simultaneous eigenvalues of  $U_f,P$ by $(u,p=\pm 1)$, it can be shown that states always appear in multiplets of the form $(u,1),(u,-1),(-u,1),(-u,-1)$ {\it i.e.}, there are two groups of degenerate states split by exactly $\pi/T$ quasienergy---hence the name $0\pi$PM. The $\pi/T$ quasienergy splitting in $\pi$SG was associated with the breaking of TTI, so it is natural to also expect TTI breaking for the $0\pi$PM. Indeed, for the special model \eqnref{eq:0 pi PMbdry}, the $\sigma^{x}$ edge operators have stroboscopic equations of motion $\sigma^{ x }_{1,N}(nT)=(-1)^n \sigma^{ x }_{1,N}(0)$, with period $2T$. At generic points in the $0\pi$PM phase obtained by perturbing \eqref{eq:0 pi PMbdry} with Ising symmetric perturbations, dressed versions of these edge Pauli operators (and generic edge operators with non-zero projections on the dressed Pauli edge operators) will exhibit period $2T$ oscillations persistent for exponentially long time scales in system size (in the same spirit as Ref.~\onlinecite{Bahri15}). Indeed, using Ising duality\cite{Khemani15}, statements about the dynamics of Ising even edge operators in the  $0\pi$PM paramagnet  directly translate into statements about local bulk operator dynamics in the (Ising symmetric) $\pi$SG in \secref{s:dynamics}. We emphasize, however, that for $0\pi$PM generic local bulk operators will not show period doubling in the limit $L\rightarrow \infty$. 

Despite the non-trivial dynamics in the $0\pi$PM, the spectral pairing properties of this phase (and the more general bosonic Floquet SPT phases discussed in Ref.~\onlinecite{vonKeyserlingkSondhi16a}) are unstable to the inclusion of small, generic symmetry breaking perturbations at the boundary. 
To see how this works in more generality, note that Floquet MBL unitaries can be re-expressed in a certain canonical form\cite{vonKeyserlingkSondhi16a}
\be\label{eq:canon}
U_{f0}=v_L v_R e^{- i f}\punc{,}
\ee
where $f$ is a local MBL Hamiltonian functional of the l-bits in the bulk, and $v_{L,R}$ are unitaries localized at the left/right edges of the system respectively which commute with the bulk l-bits. Note that the model \eqnref{eq:0 pi PMbdry} is a special realization of this more general canonical form. The SPT order of $U_{f0}$ is captured by two pieces of data: (i) The bulk SPT order, which is determined by the classification of $f$ as an undriven Hamiltonian, and (ii) the ``pumped charge'', characterized by the commutation relations between the $v_{L,R}$ and the global symmetry generators\cite{vonKeyserlingkSondhi16a}. Note that \eqnref{eq:canon} can readily be detuned -- whilst maintaining locality and unitarity -- to a form with trivial pumped charge, $e^{- i f}$,  through an interpolating family of unitaries $U_{f\lambda} = e^{-i \lambda \log v_R} e^{-i \lambda \log v_L }  U_{f 0} $ with $\lambda$ being tuned from $0$ to $1$.  Note further that if $v_{L,R}$  have non-trivial commutation relations with the global symmetry, this interpolating family of unitaries breaks the global symmetry. It may still occur that $f$, an MBL Hamiltonian, has a non-trivial  SPT classification and therefore $e^{- i f}$ has spectral pairing and edge states. However, this SPT order is readily destroyed by perturbing $f$ non-symmetrically as one would perturb an undriven SPT so as to gap out its edge states. This instability of the boundary-TTI breaking SPT phases reiterates our central message that the absolute stability of a TTI breaking phase is intrinsically tied to the coexistence of bulk spatial LRO.

The instability of $0\pi$PM SPT combined with our prior statements on the instability of pairing in the $0$SG leads to the picture depicted in \figref{Perturbed}(right)---in the presence of generic Ising symmetry breaking perturbations, the four Ising symmetric MBL-Floquet phases are reduced to two:  the absolutely stable continuation of the $\pi$SG, and a trivial PM. The $0$SG and the $0\pi$PM can be continuously connected to the trivial PM without going through a phase transition in the presence of Ising symmetry breaking terms. 

We end this section by briefly commenting on the stability of fermionic SPTs. Interacting SPTs protected by fermion parity
are more robust. Let us focus on class D\cite{Jiang11, Khemani15, vonKeyserlingkSondhi16a} for concreteness. While it is true that edge modes are unstable to fermion parity breaking perturbations, fermion parity  is never broken for physical/local Hamiltonians $H(t)$ -- hence, in the detuning argument above, $U_{f\lambda}$ is not a truly local unitary for intermediate values of $\lambda$ when $v_{L,R}$ are fermion parity odd (we say the pumped charge is fermion parity odd\cite{vonKeyserlingkSondhi16a}). However, as with all of the examples discussed here, the Floquet edge modes can be removed by breaking time translation symmetry.

\section{Concluding remarks}\label{s:conclusion}
We have shown the existence of a family of phases of Floquet systems 
which are absolutely stable---a generic interior
point in such a phase is stable to {\it all} weak local perturbations of 
its governing unitary. These phases are characterized by emergent, Hamiltonian dependent, abelian global 
symmetries and spatiotemporal long range order based on these.
Submanifolds of these phases exhibit Hamiltonian independent symmetries 
which can be unitary or anti-unitary. At
generic points in these phases, late time states evolved from randomly 
picked short ranged entangled states exhibit long range order in space
and sharp oscillations of the emergent order parameter which can be used to 
identify the phases.

These Floquet phases join two previously established paradigms for such 
absolute stability---those of topological order
and that of MBL for time independent Hamiltonians---and a comparison 
between these three is in order. Topological order, exemplified by the 
$Z_2$ order of the toric code and its weak local perturbations, is 
characterized by the absence of symmetry breaking and the presence of 
emergent gauge fields. Such phases are in a different language quantum 
liquids with long range (ground state) entanglement \cite{Chen10} which 
features account intuitively for their absolute stability.

MBL is characterized by a complete set of emergent, Hamiltonian 
dependent, local integrals of the motion (l-bits) and in its minimal 
form involves eigenstates that exhibit only short ranged entanglement. 
Its absolute stability can be attributed to the localization being 
unrelated to any spatial ordering---it is primarily a dynamical 
phenomenon. By contrast, broken symmetries are {\it not} absolutely 
stable---symmetry induced degeneracies are lifted when symmetries are 
broken.

It is not hard to believe that one can mix topological order and MBL and 
still end up with an absolutely stable phase and this was discussed as an example of eigenstate order in 
Ref.~\onlinecite{Huse13}. By contrast it is also natural to conclude
that MBL and symmetry breaking to not lead to absolute stability and 
this is also trivially the case. What is therefore
striking is that a third ingredient, Floquet periodicity, allows broken 
symmetries and MBL to combine to yield absolutely
stable phases. The resulting phases also exhibit long range entanglement 
in the form of the cat eigenstates and thus are stabilized by a relative 
of the mechanism which operates in the case of topological order.

Finally we note that the absolute stability of symmetry broken phases in this paper can be put on a similar footing to the well known absolute stability of topological phases\cite{WenNiu90}. Recent work \cite{Seiberg14,Gaiotto2015} characterizes pure abelian gauge theories as spontaneously breaking $1$-form global symmetries in their deconfined phases. In the presence of matter, the generators for these higher form symmetries are emergent and thus Hamiltonian dependent. For example, in the perturbed 2D toric code, the 1-form symmetries are generated by dressed line operators\cite{Hastings05}.  More generally, a large class of well known and undriven absolutely stable topologically ordered phases are characterized by spontaneously broken emergent $1$-form global symmetries, while the Floquet drives in this work are characterized by emergent global ($0$-form) symmetries.  In a related note, one can consider Floquet unitaries constructed from topologically ordered Hamiltonians, such as the toric code, which toggle states between different topological sectors. Such drives exhibit spatial topological order, do not break any global symmetries, but do break TTI because the Floquet unitary described does not commute with operators which measure the topological sector. Just as the cat states are split by $\pi/T$ quasi-energy in the $\pi$SG, different topological sectors are split by $\pi/T$ in this topological example. It is somewhat a matter of taste whether these should be identified as Floquet time crystals.

\acknowledgements
We thank   J. Chalker, A. Chandran, A. Lazarides, and R. Moessner for many useful discussions.  We especially thank E. Altman for making the connection to time crystals, D. Huse for his comments on dephasing, and  S. Kivelson for suggesting the term ``absolute stability''. CVK is supported by the Princeton Center for Theoretical Science. SLS and CVK  acknowledge support from the NSF-DMR via Grant No.~1311781.


\begin{thebibliography}{69}%
\makeatletter
\providecommand \@ifxundefined [1]{%
 \@ifx{#1\undefined}
}%
\providecommand \@ifnum [1]{%
 \ifnum #1\expandafter \@firstoftwo
 \else \expandafter \@secondoftwo
 \fi
}%
\providecommand \@ifx [1]{%
 \ifx #1\expandafter \@firstoftwo
 \else \expandafter \@secondoftwo
 \fi
}%
\providecommand \natexlab [1]{#1}%
\providecommand \enquote  [1]{``#1''}%
\providecommand \bibnamefont  [1]{#1}%
\providecommand \bibfnamefont [1]{#1}%
\providecommand \citenamefont [1]{#1}%
\providecommand \href@noop [0]{\@secondoftwo}%
\providecommand \href [0]{\begingroup \@sanitize@url \@href}%
\providecommand \@href[1]{\@@startlink{#1}\@@href}%
\providecommand \@@href[1]{\endgroup#1\@@endlink}%
\providecommand \@sanitize@url [0]{\catcode `\\12\catcode `\$12\catcode
  `\&12\catcode `\#12\catcode `\^12\catcode `\_12\catcode `\%12\relax}%
\providecommand \@@startlink[1]{}%
\providecommand \@@endlink[0]{}%
\providecommand \url  [0]{\begingroup\@sanitize@url \@url }%
\providecommand \@url [1]{\endgroup\@href {#1}{\urlprefix }}%
\providecommand \urlprefix  [0]{URL }%
\providecommand \Eprint [0]{\href }%
\providecommand \doibase [0]{http://dx.doi.org/}%
\providecommand \selectlanguage [0]{\@gobble}%
\providecommand \bibinfo  [0]{\@secondoftwo}%
\providecommand \bibfield  [0]{\@secondoftwo}%
\providecommand \translation [1]{[#1]}%
\providecommand \BibitemOpen [0]{}%
\providecommand \bibitemStop [0]{}%
\providecommand \bibitemNoStop [0]{.\EOS\space}%
\providecommand \EOS [0]{\spacefactor3000\relax}%
\providecommand \BibitemShut  [1]{\csname bibitem#1\endcsname}%
\let\auto@bib@innerbib\@empty
\bibitem [{\citenamefont {Chen}\ \emph {et~al.}(2010)\citenamefont {Chen},
  \citenamefont {Gu},\ and\ \citenamefont {Wen}}]{Chen10}%
  \BibitemOpen
  \bibfield  {author} {\bibinfo {author} {\bibfnamefont {X.}~\bibnamefont
  {Chen}}, \bibinfo {author} {\bibfnamefont {Z.-C.}\ \bibnamefont {Gu}}, \ and\
  \bibinfo {author} {\bibfnamefont {X.-G.}\ \bibnamefont {Wen}},\ }\href
  {\doibase 10.1103/PhysRevB.82.155138} {\bibfield  {journal} {\bibinfo
  {journal} {Phys. Rev. B}\ }\textbf {\bibinfo {volume} {82}},\ \bibinfo
  {pages} {155138} (\bibinfo {year} {2010})}\BibitemShut {NoStop}%
\bibitem [{\citenamefont {Wen}\ and\ \citenamefont {Niu}(1990)}]{WenNiu90}%
  \BibitemOpen
  \bibfield  {author} {\bibinfo {author} {\bibfnamefont {X.~G.}\ \bibnamefont
  {Wen}}\ and\ \bibinfo {author} {\bibfnamefont {Q.}~\bibnamefont {Niu}},\
  }\href {\doibase 10.1103/PhysRevB.41.9377} {\bibfield  {journal} {\bibinfo
  {journal} {Phys. Rev. B}\ }\textbf {\bibinfo {volume} {41}},\ \bibinfo
  {pages} {9377} (\bibinfo {year} {1990})}\BibitemShut {NoStop}%
\bibitem [{\citenamefont {Kitaev}(2003)}]{Kitaev03}%
  \BibitemOpen
  \bibfield  {author} {\bibinfo {author} {\bibfnamefont {A.~Y.}\ \bibnamefont
  {Kitaev}},\ }\href@noop {} {\bibfield  {journal} {\bibinfo  {journal} {Annals
  of Physics}\ }\textbf {\bibinfo {volume} {303}},\ \bibinfo {pages} {2}
  (\bibinfo {year} {2003})}\BibitemShut {NoStop}%
\bibitem [{\citenamefont {Hastings}\ and\ \citenamefont
  {Wen}(2005)}]{Hastings05}%
  \BibitemOpen
  \bibfield  {author} {\bibinfo {author} {\bibfnamefont {M.~B.}\ \bibnamefont
  {Hastings}}\ and\ \bibinfo {author} {\bibfnamefont {X.-G.}\ \bibnamefont
  {Wen}},\ }\href {\doibase 10.1103/PhysRevB.72.045141} {\bibfield  {journal}
  {\bibinfo  {journal} {Phys. Rev. B}\ }\textbf {\bibinfo {volume} {72}},\
  \bibinfo {pages} {045141} (\bibinfo {year} {2005})}\BibitemShut {NoStop}%
\bibitem [{\citenamefont {Anderson}(1958)}]{Anderson58}%
  \BibitemOpen
  \bibfield  {author} {\bibinfo {author} {\bibfnamefont {P.~W.}\ \bibnamefont
  {Anderson}},\ }\href {\doibase 10.1103/PhysRev.109.1492} {\bibfield
  {journal} {\bibinfo  {journal} {Phys. Rev.}\ }\textbf {\bibinfo {volume}
  {109}},\ \bibinfo {pages} {1492} (\bibinfo {year} {1958})}\BibitemShut
  {NoStop}%
\bibitem [{\citenamefont {Basko}\ \emph {et~al.}(2006)\citenamefont {Basko},
  \citenamefont {Aleiner},\ and\ \citenamefont {Altshuler}}]{Basko06}%
  \BibitemOpen
  \bibfield  {author} {\bibinfo {author} {\bibfnamefont {D.~M.}\ \bibnamefont
  {Basko}}, \bibinfo {author} {\bibfnamefont {I.~L.}\ \bibnamefont {Aleiner}},
  \ and\ \bibinfo {author} {\bibfnamefont {B.~L.}\ \bibnamefont {Altshuler}},\
  }\href {\doibase 10.1016/j.aop.2005.11.014} {\bibfield  {journal} {\bibinfo
  {journal} {Annals of Physics}\ }\textbf {\bibinfo {volume} {321}},\ \bibinfo
  {pages} {1126} (\bibinfo {year} {2006})}\BibitemShut {NoStop}%
\bibitem [{\citenamefont {Pal}\ and\ \citenamefont {Huse}(2010)}]{PalHuse}%
  \BibitemOpen
  \bibfield  {author} {\bibinfo {author} {\bibfnamefont {A.}~\bibnamefont
  {Pal}}\ and\ \bibinfo {author} {\bibfnamefont {D.~A.}\ \bibnamefont {Huse}},\
  }\href {\doibase 10.1103/PhysRevB.82.174411} {\bibfield  {journal} {\bibinfo
  {journal} {Phys. Rev. B}\ }\textbf {\bibinfo {volume} {82}},\ \bibinfo
  {pages} {174411} (\bibinfo {year} {2010})}\BibitemShut {NoStop}%
\bibitem [{\citenamefont {Oganesyan}\ and\ \citenamefont
  {Huse}(2007)}]{OganesyanHuse}%
  \BibitemOpen
  \bibfield  {author} {\bibinfo {author} {\bibfnamefont {V.}~\bibnamefont
  {Oganesyan}}\ and\ \bibinfo {author} {\bibfnamefont {D.~A.}\ \bibnamefont
  {Huse}},\ }\href {\doibase 10.1103/PhysRevB.75.155111} {\bibfield  {journal}
  {\bibinfo  {journal} {Phys. Rev. B}\ }\textbf {\bibinfo {volume} {75}},\
  \bibinfo {pages} {155111} (\bibinfo {year} {2007})}\BibitemShut {NoStop}%
\bibitem [{\citenamefont {{Nandkishore}}\ and\ \citenamefont
  {{Huse}}(2015)}]{Nandkishore14}%
  \BibitemOpen
  \bibfield  {author} {\bibinfo {author} {\bibfnamefont {R.}~\bibnamefont
  {{Nandkishore}}}\ and\ \bibinfo {author} {\bibfnamefont {D.~A.}\ \bibnamefont
  {{Huse}}},\ }\href {\doibase 10.1146/annurev-conmatphys-031214-014726}
  {\bibfield  {journal} {\bibinfo  {journal} {Annual Review of Condensed Matter
  Physics}\ }\textbf {\bibinfo {volume} {6}},\ \bibinfo {pages} {15} (\bibinfo
  {year} {2015})},\ \Eprint {http://arxiv.org/abs/1404.0686} {arXiv:1404.0686
  [cond-mat.stat-mech]} \BibitemShut {NoStop}%
\bibitem [{\citenamefont {Altman}\ and\ \citenamefont
  {Vosk}(2015)}]{AltmanVosk}%
  \BibitemOpen
  \bibfield  {author} {\bibinfo {author} {\bibfnamefont {E.}~\bibnamefont
  {Altman}}\ and\ \bibinfo {author} {\bibfnamefont {R.}~\bibnamefont {Vosk}},\
  }\href {\doibase 10.1146/annurev-conmatphys-031214-014701} {\bibfield
  {journal} {\bibinfo  {journal} {Annual Review of Condensed Matter Physics}\
  }\textbf {\bibinfo {volume} {6}},\ \bibinfo {pages} {383} (\bibinfo {year}
  {2015})}\BibitemShut {NoStop}%
\bibitem [{\citenamefont {Huse}\ \emph {et~al.}(2014)\citenamefont {Huse},
  \citenamefont {Nandkishore},\ and\ \citenamefont {Oganesyan}}]{Huse14}%
  \BibitemOpen
  \bibfield  {author} {\bibinfo {author} {\bibfnamefont {D.~A.}\ \bibnamefont
  {Huse}}, \bibinfo {author} {\bibfnamefont {R.}~\bibnamefont {Nandkishore}}, \
  and\ \bibinfo {author} {\bibfnamefont {V.}~\bibnamefont {Oganesyan}},\ }\href
  {\doibase 10.1103/PhysRevB.90.174202} {\bibfield  {journal} {\bibinfo
  {journal} {Phys. Rev. B}\ }\textbf {\bibinfo {volume} {90}},\ \bibinfo
  {pages} {174202} (\bibinfo {year} {2014})}\BibitemShut {NoStop}%
\bibitem [{\citenamefont {Serbyn}\ \emph
  {et~al.}(2013{\natexlab{a}})\citenamefont {Serbyn}, \citenamefont
  {Papi\ifmmode~\acute{c}\else \'{c}\fi{}},\ and\ \citenamefont
  {Abanin}}]{Serbyn13cons}%
  \BibitemOpen
  \bibfield  {author} {\bibinfo {author} {\bibfnamefont {M.}~\bibnamefont
  {Serbyn}}, \bibinfo {author} {\bibfnamefont {Z.}~\bibnamefont
  {Papi\ifmmode~\acute{c}\else \'{c}\fi{}}}, \ and\ \bibinfo {author}
  {\bibfnamefont {D.~A.}\ \bibnamefont {Abanin}},\ }\href {\doibase
  10.1103/PhysRevLett.111.127201} {\bibfield  {journal} {\bibinfo  {journal}
  {Phys. Rev. Lett.}\ }\textbf {\bibinfo {volume} {111}},\ \bibinfo {pages}
  {127201} (\bibinfo {year} {2013}{\natexlab{a}})}\BibitemShut {NoStop}%
\bibitem [{\citenamefont {Imbrie}(2016)}]{Imbrie2016}%
  \BibitemOpen
  \bibfield  {author} {\bibinfo {author} {\bibfnamefont {J.~Z.}\ \bibnamefont
  {Imbrie}},\ }\href {\doibase 10.1007/s10955-016-1508-x} {\bibfield  {journal}
  {\bibinfo  {journal} {Journal of Statistical Physics}\ }\textbf {\bibinfo
  {volume} {163}},\ \bibinfo {pages} {998} (\bibinfo {year}
  {2016})}\BibitemShut {NoStop}%
\bibitem [{\citenamefont {Serbyn}\ \emph
  {et~al.}(2013{\natexlab{b}})\citenamefont {Serbyn}, \citenamefont
  {Papi\ifmmode~\acute{c}\else \'{c}\fi{}},\ and\ \citenamefont
  {Abanin}}]{Serbyn13a}%
  \BibitemOpen
  \bibfield  {author} {\bibinfo {author} {\bibfnamefont {M.}~\bibnamefont
  {Serbyn}}, \bibinfo {author} {\bibfnamefont {Z.}~\bibnamefont
  {Papi\ifmmode~\acute{c}\else \'{c}\fi{}}}, \ and\ \bibinfo {author}
  {\bibfnamefont {D.~A.}\ \bibnamefont {Abanin}},\ }\href {\doibase
  10.1103/PhysRevLett.110.260601} {\bibfield  {journal} {\bibinfo  {journal}
  {Phys. Rev. Lett.}\ }\textbf {\bibinfo {volume} {110}},\ \bibinfo {pages}
  {260601} (\bibinfo {year} {2013}{\natexlab{b}})}\BibitemShut {NoStop}%
\bibitem [{\citenamefont {Chandran}\ \emph {et~al.}(2015)\citenamefont
  {Chandran}, \citenamefont {Kim}, \citenamefont {Vidal},\ and\ \citenamefont
  {Abanin}}]{Chandran15b}%
  \BibitemOpen
  \bibfield  {author} {\bibinfo {author} {\bibfnamefont {A.}~\bibnamefont
  {Chandran}}, \bibinfo {author} {\bibfnamefont {I.~H.}\ \bibnamefont {Kim}},
  \bibinfo {author} {\bibfnamefont {G.}~\bibnamefont {Vidal}}, \ and\ \bibinfo
  {author} {\bibfnamefont {D.~A.}\ \bibnamefont {Abanin}},\ }\href {\doibase
  10.1103/PhysRevB.91.085425} {\bibfield  {journal} {\bibinfo  {journal} {Phys.
  Rev. B}\ }\textbf {\bibinfo {volume} {91}},\ \bibinfo {pages} {085425}
  (\bibinfo {year} {2015})}\BibitemShut {NoStop}%
\bibitem [{\citenamefont {{Ros}}\ \emph {et~al.}(2015)\citenamefont {{Ros}},
  \citenamefont {{M{\"u}ller}},\ and\ \citenamefont {{Scardicchio}}}]{Ros15}%
  \BibitemOpen
  \bibfield  {author} {\bibinfo {author} {\bibfnamefont {V.}~\bibnamefont
  {{Ros}}}, \bibinfo {author} {\bibfnamefont {M.}~\bibnamefont {{M{\"u}ller}}},
  \ and\ \bibinfo {author} {\bibfnamefont {A.}~\bibnamefont {{Scardicchio}}},\
  }\href {\doibase 10.1016/j.nuclphysb.2014.12.014} {\bibfield  {journal}
  {\bibinfo  {journal} {Nuclear Physics B}\ }\textbf {\bibinfo {volume}
  {891}},\ \bibinfo {pages} {420} (\bibinfo {year} {2015})},\ \Eprint
  {http://arxiv.org/abs/1406.2175} {arXiv:1406.2175 [cond-mat.dis-nn]}
  \BibitemShut {NoStop}%
\bibitem [{\citenamefont {{Pekker}}\ and\ \citenamefont
  {{Clark}}(2014)}]{Pekker14}%
  \BibitemOpen
  \bibfield  {author} {\bibinfo {author} {\bibfnamefont {D.}~\bibnamefont
  {{Pekker}}}\ and\ \bibinfo {author} {\bibfnamefont {B.~K.}\ \bibnamefont
  {{Clark}}},\ }\href@noop {} {\bibfield  {journal} {\bibinfo  {journal} {ArXiv
  e-prints}\ } (\bibinfo {year} {2014})},\ \Eprint
  {http://arxiv.org/abs/1410.2224} {arXiv:1410.2224 [cond-mat.str-el]}
  \BibitemShut {NoStop}%
\bibitem [{\citenamefont {Rademaker}\ and\ \citenamefont
  {Ortu\~no}(2016)}]{Rademaker2016}%
  \BibitemOpen
  \bibfield  {author} {\bibinfo {author} {\bibfnamefont {L.}~\bibnamefont
  {Rademaker}}\ and\ \bibinfo {author} {\bibfnamefont {M.}~\bibnamefont
  {Ortu\~no}},\ }\href {\doibase 10.1103/PhysRevLett.116.010404} {\bibfield
  {journal} {\bibinfo  {journal} {Phys. Rev. Lett.}\ }\textbf {\bibinfo
  {volume} {116}},\ \bibinfo {pages} {010404} (\bibinfo {year}
  {2016})}\BibitemShut {NoStop}%
\bibitem [{\citenamefont {Huse}\ \emph {et~al.}(2013)\citenamefont {Huse},
  \citenamefont {Nandkishore}, \citenamefont {Oganesyan}, \citenamefont {Pal},\
  and\ \citenamefont {Sondhi}}]{Huse13}%
  \BibitemOpen
  \bibfield  {author} {\bibinfo {author} {\bibfnamefont {D.~A.}\ \bibnamefont
  {Huse}}, \bibinfo {author} {\bibfnamefont {R.}~\bibnamefont {Nandkishore}},
  \bibinfo {author} {\bibfnamefont {V.}~\bibnamefont {Oganesyan}}, \bibinfo
  {author} {\bibfnamefont {A.}~\bibnamefont {Pal}}, \ and\ \bibinfo {author}
  {\bibfnamefont {S.~L.}\ \bibnamefont {Sondhi}},\ }\href {\doibase
  10.1103/PhysRevB.88.014206} {\bibfield  {journal} {\bibinfo  {journal} {Phys.
  Rev. B}\ }\textbf {\bibinfo {volume} {88}},\ \bibinfo {pages} {014206}
  (\bibinfo {year} {2013})}\BibitemShut {NoStop}%
\bibitem [{\citenamefont {Pekker}\ \emph {et~al.}(2014)\citenamefont {Pekker},
  \citenamefont {Refael}, \citenamefont {Altman}, \citenamefont {Demler},\ and\
  \citenamefont {Oganesyan}}]{PekkerHilbertGlass}%
  \BibitemOpen
  \bibfield  {author} {\bibinfo {author} {\bibfnamefont {D.}~\bibnamefont
  {Pekker}}, \bibinfo {author} {\bibfnamefont {G.}~\bibnamefont {Refael}},
  \bibinfo {author} {\bibfnamefont {E.}~\bibnamefont {Altman}}, \bibinfo
  {author} {\bibfnamefont {E.}~\bibnamefont {Demler}}, \ and\ \bibinfo {author}
  {\bibfnamefont {V.}~\bibnamefont {Oganesyan}},\ }\href {\doibase
  10.1103/PhysRevX.4.011052} {\bibfield  {journal} {\bibinfo  {journal} {Phys.
  Rev. X}\ }\textbf {\bibinfo {volume} {4}},\ \bibinfo {pages} {011052}
  (\bibinfo {year} {2014})}\BibitemShut {NoStop}%
\bibitem [{\citenamefont {Bauer}\ and\ \citenamefont {Nayak}(2013)}]{Bauer13}%
  \BibitemOpen
  \bibfield  {author} {\bibinfo {author} {\bibfnamefont {B.}~\bibnamefont
  {Bauer}}\ and\ \bibinfo {author} {\bibfnamefont {C.}~\bibnamefont {Nayak}},\
  }\href {http://stacks.iop.org/1742-5468/2013/i=09/a=P09005} {\bibfield
  {journal} {\bibinfo  {journal} {Journal of Statistical Mechanics: Theory and
  Experiment}\ }\textbf {\bibinfo {volume} {2013}},\ \bibinfo {pages} {P09005}
  (\bibinfo {year} {2013})}\BibitemShut {NoStop}%
\bibitem [{\citenamefont {Chandran}\ \emph {et~al.}(2014)\citenamefont
  {Chandran}, \citenamefont {Khemani}, \citenamefont {Laumann},\ and\
  \citenamefont {Sondhi}}]{Chandran14}%
  \BibitemOpen
  \bibfield  {author} {\bibinfo {author} {\bibfnamefont {A.}~\bibnamefont
  {Chandran}}, \bibinfo {author} {\bibfnamefont {V.}~\bibnamefont {Khemani}},
  \bibinfo {author} {\bibfnamefont {C.~R.}\ \bibnamefont {Laumann}}, \ and\
  \bibinfo {author} {\bibfnamefont {S.~L.}\ \bibnamefont {Sondhi}},\ }\href
  {\doibase 10.1103/PhysRevB.89.144201} {\bibfield  {journal} {\bibinfo
  {journal} {Phys. Rev. B}\ }\textbf {\bibinfo {volume} {89}},\ \bibinfo
  {pages} {144201} (\bibinfo {year} {2014})}\BibitemShut {NoStop}%
\bibitem [{\citenamefont {Bahri}\ \emph {et~al.}(2015)\citenamefont {Bahri},
  \citenamefont {Vosk}, \citenamefont {Altman},\ and\ \citenamefont
  {Vishwanath}}]{Bahri15}%
  \BibitemOpen
  \bibfield  {author} {\bibinfo {author} {\bibfnamefont {Y.}~\bibnamefont
  {Bahri}}, \bibinfo {author} {\bibfnamefont {R.}~\bibnamefont {Vosk}},
  \bibinfo {author} {\bibfnamefont {E.}~\bibnamefont {Altman}}, \ and\ \bibinfo
  {author} {\bibfnamefont {A.}~\bibnamefont {Vishwanath}},\ }\href
  {http://dx.doi.org/10.1038/ncomms8341} {\bibfield  {journal} {\bibinfo
  {journal} {Nat Commun}\ }\textbf {\bibinfo {volume} {6}} (\bibinfo {year}
  {2015})}\BibitemShut {NoStop}%
\bibitem [{\citenamefont {Potter}\ and\ \citenamefont
  {Vishwanath}(2015)}]{Potter15}%
  \BibitemOpen
  \bibfield  {author} {\bibinfo {author} {\bibfnamefont {A.~C.}\ \bibnamefont
  {Potter}}\ and\ \bibinfo {author} {\bibfnamefont {A.}~\bibnamefont
  {Vishwanath}},\ }\href@noop {} {\bibfield  {journal} {\bibinfo  {journal}
  {arXiv preprint arXiv:1506.00592}\ } (\bibinfo {year} {2015})}\BibitemShut
  {NoStop}%
\bibitem [{\citenamefont {Khemani}\ \emph {et~al.}(2015)\citenamefont
  {Khemani}, \citenamefont {Lazarides}, \citenamefont {Moessner},\ and\
  \citenamefont {Sondhi}}]{Khemani15}%
  \BibitemOpen
  \bibfield  {author} {\bibinfo {author} {\bibfnamefont {V.}~\bibnamefont
  {Khemani}}, \bibinfo {author} {\bibfnamefont {A.}~\bibnamefont {Lazarides}},
  \bibinfo {author} {\bibfnamefont {R.}~\bibnamefont {Moessner}}, \ and\
  \bibinfo {author} {\bibfnamefont {S.~L.}\ \bibnamefont {Sondhi}},\
  }\href@noop {} {\bibfield  {journal} {\bibinfo  {journal} {arXiv preprint
  arXiv:1508.03344}\ } (\bibinfo {year} {2015})}\BibitemShut {NoStop}%
\bibitem [{\citenamefont {{von Keyserlingk}}\ and\ \citenamefont
  {{Sondhi}}(2016{\natexlab{a}})}]{vonKeyserlingkSondhi16b}%
  \BibitemOpen
  \bibfield  {author} {\bibinfo {author} {\bibfnamefont {C.~W.}\ \bibnamefont
  {{von Keyserlingk}}}\ and\ \bibinfo {author} {\bibfnamefont {S.~L.}\
  \bibnamefont {{Sondhi}}},\ }\href@noop {} {\bibfield  {journal} {\bibinfo
  {journal} {ArXiv e-prints}\ } (\bibinfo {year} {2016}{\natexlab{a}})},\
  \Eprint {http://arxiv.org/abs/1602.06949} {arXiv:1602.06949
  [cond-mat.str-el]} \BibitemShut {NoStop}%
\bibitem [{\citenamefont {Kitagawa}\ \emph {et~al.}(2010)\citenamefont
  {Kitagawa}, \citenamefont {Berg}, \citenamefont {Rudner},\ and\ \citenamefont
  {Demler}}]{Kitagawa10}%
  \BibitemOpen
  \bibfield  {author} {\bibinfo {author} {\bibfnamefont {T.}~\bibnamefont
  {Kitagawa}}, \bibinfo {author} {\bibfnamefont {E.}~\bibnamefont {Berg}},
  \bibinfo {author} {\bibfnamefont {M.}~\bibnamefont {Rudner}}, \ and\ \bibinfo
  {author} {\bibfnamefont {E.}~\bibnamefont {Demler}},\ }\href {\doibase
  10.1103/PhysRevB.82.235114} {\bibfield  {journal} {\bibinfo  {journal} {Phys.
  Rev. B}\ }\textbf {\bibinfo {volume} {82}},\ \bibinfo {pages} {235114}
  (\bibinfo {year} {2010})}\BibitemShut {NoStop}%
\bibitem [{\citenamefont {Jiang}\ \emph {et~al.}(2011)\citenamefont {Jiang},
  \citenamefont {Kitagawa}, \citenamefont {Alicea}, \citenamefont {Akhmerov},
  \citenamefont {Pekker}, \citenamefont {Refael}, \citenamefont {Cirac},
  \citenamefont {Demler}, \citenamefont {Lukin},\ and\ \citenamefont
  {Zoller}}]{Jiang11}%
  \BibitemOpen
  \bibfield  {author} {\bibinfo {author} {\bibfnamefont {L.}~\bibnamefont
  {Jiang}}, \bibinfo {author} {\bibfnamefont {T.}~\bibnamefont {Kitagawa}},
  \bibinfo {author} {\bibfnamefont {J.}~\bibnamefont {Alicea}}, \bibinfo
  {author} {\bibfnamefont {A.~R.}\ \bibnamefont {Akhmerov}}, \bibinfo {author}
  {\bibfnamefont {D.}~\bibnamefont {Pekker}}, \bibinfo {author} {\bibfnamefont
  {G.}~\bibnamefont {Refael}}, \bibinfo {author} {\bibfnamefont {J.~I.}\
  \bibnamefont {Cirac}}, \bibinfo {author} {\bibfnamefont {E.}~\bibnamefont
  {Demler}}, \bibinfo {author} {\bibfnamefont {M.~D.}\ \bibnamefont {Lukin}}, \
  and\ \bibinfo {author} {\bibfnamefont {P.}~\bibnamefont {Zoller}},\ }\href
  {\doibase 10.1103/PhysRevLett.106.220402} {\bibfield  {journal} {\bibinfo
  {journal} {Phys. Rev. Lett.}\ }\textbf {\bibinfo {volume} {106}},\ \bibinfo
  {pages} {220402} (\bibinfo {year} {2011})}\BibitemShut {NoStop}%
\bibitem [{\citenamefont {Lindner}\ \emph {et~al.}(2011)\citenamefont
  {Lindner}, \citenamefont {Refael},\ and\ \citenamefont
  {Galitski}}]{Lindner11}%
  \BibitemOpen
  \bibfield  {author} {\bibinfo {author} {\bibfnamefont {N.~H.}\ \bibnamefont
  {Lindner}}, \bibinfo {author} {\bibfnamefont {G.}~\bibnamefont {Refael}}, \
  and\ \bibinfo {author} {\bibfnamefont {V.}~\bibnamefont {Galitski}},\ }\href
  {\doibase 10.1038/nphys1926} {\bibfield  {journal} {\bibinfo  {journal}
  {Nature Physics}\ }\textbf {\bibinfo {volume} {7}},\ \bibinfo {pages} {490}
  (\bibinfo {year} {2011})}\BibitemShut {NoStop}%
\bibitem [{\citenamefont {Thakurathi}\ \emph {et~al.}(2013)\citenamefont
  {Thakurathi}, \citenamefont {Patel}, \citenamefont {Sen},\ and\ \citenamefont
  {Dutta}}]{Thakurathi13}%
  \BibitemOpen
  \bibfield  {author} {\bibinfo {author} {\bibfnamefont {M.}~\bibnamefont
  {Thakurathi}}, \bibinfo {author} {\bibfnamefont {A.~A.}\ \bibnamefont
  {Patel}}, \bibinfo {author} {\bibfnamefont {D.}~\bibnamefont {Sen}}, \ and\
  \bibinfo {author} {\bibfnamefont {A.}~\bibnamefont {Dutta}},\ }\href
  {\doibase 10.1103/PhysRevB.88.155133} {\bibfield  {journal} {\bibinfo
  {journal} {Phys. Rev. B}\ }\textbf {\bibinfo {volume} {88}},\ \bibinfo
  {pages} {155133} (\bibinfo {year} {2013})}\BibitemShut {NoStop}%
\bibitem [{\citenamefont {Rudner}\ \emph {et~al.}(2013)\citenamefont {Rudner},
  \citenamefont {Lindner}, \citenamefont {Berg},\ and\ \citenamefont
  {Levin}}]{Rudner13}%
  \BibitemOpen
  \bibfield  {author} {\bibinfo {author} {\bibfnamefont {M.~S.}\ \bibnamefont
  {Rudner}}, \bibinfo {author} {\bibfnamefont {N.~H.}\ \bibnamefont {Lindner}},
  \bibinfo {author} {\bibfnamefont {E.}~\bibnamefont {Berg}}, \ and\ \bibinfo
  {author} {\bibfnamefont {M.}~\bibnamefont {Levin}},\ }\href {\doibase
  10.1103/PhysRevX.3.031005} {\bibfield  {journal} {\bibinfo  {journal} {Phys.
  Rev. X}\ }\textbf {\bibinfo {volume} {3}},\ \bibinfo {pages} {031005}
  (\bibinfo {year} {2013})}\BibitemShut {NoStop}%
\bibitem [{\citenamefont {Asb\'oth}\ \emph {et~al.}(2014)\citenamefont
  {Asb\'oth}, \citenamefont {Tarasinski},\ and\ \citenamefont
  {Delplace}}]{Asboth14}%
  \BibitemOpen
  \bibfield  {author} {\bibinfo {author} {\bibfnamefont {J.~K.}\ \bibnamefont
  {Asb\'oth}}, \bibinfo {author} {\bibfnamefont {B.}~\bibnamefont
  {Tarasinski}}, \ and\ \bibinfo {author} {\bibfnamefont {P.}~\bibnamefont
  {Delplace}},\ }\href {\doibase 10.1103/PhysRevB.90.125143} {\bibfield
  {journal} {\bibinfo  {journal} {Phys. Rev. B}\ }\textbf {\bibinfo {volume}
  {90}},\ \bibinfo {pages} {125143} (\bibinfo {year} {2014})}\BibitemShut
  {NoStop}%
\bibitem [{\citenamefont {Carpentier}\ \emph {et~al.}(2015)\citenamefont
  {Carpentier}, \citenamefont {Delplace}, \citenamefont {Fruchart},\ and\
  \citenamefont {Gawedzki}}]{Carpentier15}%
  \BibitemOpen
  \bibfield  {author} {\bibinfo {author} {\bibfnamefont {D.}~\bibnamefont
  {Carpentier}}, \bibinfo {author} {\bibfnamefont {P.}~\bibnamefont
  {Delplace}}, \bibinfo {author} {\bibfnamefont {M.}~\bibnamefont {Fruchart}},
  \ and\ \bibinfo {author} {\bibfnamefont {K.}~\bibnamefont {Gawedzki}},\
  }\href {\doibase 10.1103/PhysRevLett.114.106806} {\bibfield  {journal}
  {\bibinfo  {journal} {Phys. Rev. Lett.}\ }\textbf {\bibinfo {volume} {114}},\
  \bibinfo {pages} {106806} (\bibinfo {year} {2015})}\BibitemShut {NoStop}%
\bibitem [{\citenamefont {Nathan}\ and\ \citenamefont
  {Rudner}(2015)}]{Nathan15}%
  \BibitemOpen
  \bibfield  {author} {\bibinfo {author} {\bibfnamefont {F.}~\bibnamefont
  {Nathan}}\ and\ \bibinfo {author} {\bibfnamefont {M.~S.}\ \bibnamefont
  {Rudner}},\ }\href {http://stacks.iop.org/1367-2630/17/i=12/a=125014}
  {\bibfield  {journal} {\bibinfo  {journal} {New Journal of Physics}\ }\textbf
  {\bibinfo {volume} {17}},\ \bibinfo {pages} {125014} (\bibinfo {year}
  {2015})}\BibitemShut {NoStop}%
\bibitem [{\citenamefont {{Roy}}\ and\ \citenamefont
  {{Harper}}(2016{\natexlab{a}})}]{Roy15}%
  \BibitemOpen
  \bibfield  {author} {\bibinfo {author} {\bibfnamefont {R.}~\bibnamefont
  {{Roy}}}\ and\ \bibinfo {author} {\bibfnamefont {F.}~\bibnamefont
  {{Harper}}},\ }\href@noop {} {\bibfield  {journal} {\bibinfo  {journal}
  {ArXiv e-prints}\ } (\bibinfo {year} {2016}{\natexlab{a}})},\ \Eprint
  {http://arxiv.org/abs/1603.06944} {arXiv:1603.06944 [cond-mat.str-el]}
  \BibitemShut {NoStop}%
\bibitem [{\citenamefont {Titum}\ \emph
  {et~al.}(2015{\natexlab{a}})\citenamefont {Titum}, \citenamefont {Lindner},
  \citenamefont {Rechtsman},\ and\ \citenamefont {Refael}}]{Titum15a}%
  \BibitemOpen
  \bibfield  {author} {\bibinfo {author} {\bibfnamefont {P.}~\bibnamefont
  {Titum}}, \bibinfo {author} {\bibfnamefont {N.~H.}\ \bibnamefont {Lindner}},
  \bibinfo {author} {\bibfnamefont {M.~C.}\ \bibnamefont {Rechtsman}}, \ and\
  \bibinfo {author} {\bibfnamefont {G.}~\bibnamefont {Refael}},\ }\href
  {\doibase 10.1103/PhysRevLett.114.056801} {\bibfield  {journal} {\bibinfo
  {journal} {Phys. Rev. Lett.}\ }\textbf {\bibinfo {volume} {114}},\ \bibinfo
  {pages} {056801} (\bibinfo {year} {2015}{\natexlab{a}})}\BibitemShut
  {NoStop}%
\bibitem [{\citenamefont {Titum}\ \emph
  {et~al.}(2015{\natexlab{b}})\citenamefont {Titum}, \citenamefont {Berg},
  \citenamefont {Rudner}, \citenamefont {Refael},\ and\ \citenamefont
  {Lindner}}]{Titum15b}%
  \BibitemOpen
  \bibfield  {author} {\bibinfo {author} {\bibfnamefont {P.}~\bibnamefont
  {Titum}}, \bibinfo {author} {\bibfnamefont {E.}~\bibnamefont {Berg}},
  \bibinfo {author} {\bibfnamefont {M.~S.}\ \bibnamefont {Rudner}}, \bibinfo
  {author} {\bibfnamefont {G.}~\bibnamefont {Refael}}, \ and\ \bibinfo {author}
  {\bibfnamefont {N.~H.}\ \bibnamefont {Lindner}},\ }\href@noop {} {\bibfield
  {journal} {\bibinfo  {journal} {arXiv preprint arXiv:1506.00650}\ } (\bibinfo
  {year} {2015}{\natexlab{b}})}\BibitemShut {NoStop}%
\bibitem [{\citenamefont {Lazarides}\ \emph {et~al.}(2015)\citenamefont
  {Lazarides}, \citenamefont {Das},\ and\ \citenamefont
  {Moessner}}]{Lazarides14}%
  \BibitemOpen
  \bibfield  {author} {\bibinfo {author} {\bibfnamefont {A.}~\bibnamefont
  {Lazarides}}, \bibinfo {author} {\bibfnamefont {A.}~\bibnamefont {Das}}, \
  and\ \bibinfo {author} {\bibfnamefont {R.}~\bibnamefont {Moessner}},\ }\href
  {\doibase 10.1103/PhysRevLett.115.030402} {\bibfield  {journal} {\bibinfo
  {journal} {Phys. Rev. Lett.}\ }\textbf {\bibinfo {volume} {115}},\ \bibinfo
  {pages} {030402} (\bibinfo {year} {2015})}\BibitemShut {NoStop}%
\bibitem [{\citenamefont {Ponte}\ \emph
  {et~al.}(2015{\natexlab{a}})\citenamefont {Ponte}, \citenamefont
  {Papi\ifmmode~\acute{c}\else \'{c}\fi{}}, \citenamefont {Huveneers},\ and\
  \citenamefont {Abanin}}]{Ponte15}%
  \BibitemOpen
  \bibfield  {author} {\bibinfo {author} {\bibfnamefont {P.}~\bibnamefont
  {Ponte}}, \bibinfo {author} {\bibfnamefont {Z.}~\bibnamefont
  {Papi\ifmmode~\acute{c}\else \'{c}\fi{}}}, \bibinfo {author} {\bibfnamefont
  {F.}~\bibnamefont {Huveneers}}, \ and\ \bibinfo {author} {\bibfnamefont
  {D.~A.}\ \bibnamefont {Abanin}},\ }\href {\doibase
  10.1103/PhysRevLett.114.140401} {\bibfield  {journal} {\bibinfo  {journal}
  {Phys. Rev. Lett.}\ }\textbf {\bibinfo {volume} {114}},\ \bibinfo {pages}
  {140401} (\bibinfo {year} {2015}{\natexlab{a}})}\BibitemShut {NoStop}%
\bibitem [{\citenamefont {Ponte}\ \emph
  {et~al.}(2015{\natexlab{b}})\citenamefont {Ponte}, \citenamefont {Chandran},
  \citenamefont {Papić},\ and\ \citenamefont {Abanin}}]{Ponte15b}%
  \BibitemOpen
  \bibfield  {author} {\bibinfo {author} {\bibfnamefont {P.}~\bibnamefont
  {Ponte}}, \bibinfo {author} {\bibfnamefont {A.}~\bibnamefont {Chandran}},
  \bibinfo {author} {\bibfnamefont {Z.}~\bibnamefont {Papić}}, \ and\ \bibinfo
  {author} {\bibfnamefont {D.~A.}\ \bibnamefont {Abanin}},\ }\href {\doibase
  http://dx.doi.org/10.1016/j.aop.2014.11.008} {\bibfield  {journal} {\bibinfo
  {journal} {Annals of Physics}\ }\textbf {\bibinfo {volume} {353}},\ \bibinfo
  {pages} {196 } (\bibinfo {year} {2015}{\natexlab{b}})}\BibitemShut {NoStop}%
\bibitem [{\citenamefont {Abanin}\ \emph {et~al.}(2014)\citenamefont {Abanin},
  \citenamefont {De~Roeck},\ and\ \citenamefont {Huveneers}}]{Abanin14}%
  \BibitemOpen
  \bibfield  {author} {\bibinfo {author} {\bibfnamefont {D.}~\bibnamefont
  {Abanin}}, \bibinfo {author} {\bibfnamefont {W.}~\bibnamefont {De~Roeck}}, \
  and\ \bibinfo {author} {\bibfnamefont {F.}~\bibnamefont {Huveneers}},\
  }\href@noop {} {\bibfield  {journal} {\bibinfo  {journal} {arXiv preprint
  arXiv:1412.4752}\ } (\bibinfo {year} {2014})}\BibitemShut {NoStop}%
\bibitem [{\citenamefont {D'Alessio}\ and\ \citenamefont
  {Rigol}(2014)}]{Rigol14}%
  \BibitemOpen
  \bibfield  {author} {\bibinfo {author} {\bibfnamefont {L.}~\bibnamefont
  {D'Alessio}}\ and\ \bibinfo {author} {\bibfnamefont {M.}~\bibnamefont
  {Rigol}},\ }\href {\doibase 10.1103/PhysRevX.4.041048} {\bibfield  {journal}
  {\bibinfo  {journal} {Phys. Rev. X}\ }\textbf {\bibinfo {volume} {4}},\
  \bibinfo {pages} {041048} (\bibinfo {year} {2014})}\BibitemShut {NoStop}%
\bibitem [{\citenamefont {{von Keyserlingk}}\ and\ \citenamefont
  {{Sondhi}}(2016{\natexlab{b}})}]{vonKeyserlingkSondhi16a}%
  \BibitemOpen
  \bibfield  {author} {\bibinfo {author} {\bibfnamefont {C.~W.}\ \bibnamefont
  {{von Keyserlingk}}}\ and\ \bibinfo {author} {\bibfnamefont {S.~L.}\
  \bibnamefont {{Sondhi}}},\ }\href@noop {} {\bibfield  {journal} {\bibinfo
  {journal} {ArXiv e-prints}\ } (\bibinfo {year} {2016}{\natexlab{b}})},\
  \Eprint {http://arxiv.org/abs/1602.02157} {arXiv:1602.02157
  [cond-mat.str-el]} \BibitemShut {NoStop}%
\bibitem [{\citenamefont {{Else}}\ and\ \citenamefont
  {{Nayak}}(2016)}]{Else16}%
  \BibitemOpen
  \bibfield  {author} {\bibinfo {author} {\bibfnamefont {D.~V.}\ \bibnamefont
  {{Else}}}\ and\ \bibinfo {author} {\bibfnamefont {C.}~\bibnamefont
  {{Nayak}}},\ }\href@noop {} {\bibfield  {journal} {\bibinfo  {journal} {ArXiv
  e-prints}\ } (\bibinfo {year} {2016})},\ \Eprint
  {http://arxiv.org/abs/1602.04804} {arXiv:1602.04804 [cond-mat.str-el]}
  \BibitemShut {NoStop}%
\bibitem [{\citenamefont {{Potter}}\ \emph {et~al.}(2016)\citenamefont
  {{Potter}}, \citenamefont {{Morimoto}},\ and\ \citenamefont
  {{Vishwanath}}}]{Potter16}%
  \BibitemOpen
  \bibfield  {author} {\bibinfo {author} {\bibfnamefont {A.~C.}\ \bibnamefont
  {{Potter}}}, \bibinfo {author} {\bibfnamefont {T.}~\bibnamefont
  {{Morimoto}}}, \ and\ \bibinfo {author} {\bibfnamefont {A.}~\bibnamefont
  {{Vishwanath}}},\ }\href@noop {} {\bibfield  {journal} {\bibinfo  {journal}
  {ArXiv e-prints}\ } (\bibinfo {year} {2016})},\ \Eprint
  {http://arxiv.org/abs/1602.05194} {arXiv:1602.05194 [cond-mat.str-el]}
  \BibitemShut {NoStop}%
\bibitem [{Note1()}]{Note1}%
  \BibitemOpen
  \bibinfo {note} {We use the term ``spontaneously break unitary global
  symmetries'' to mean that the eigenstates exhibit the long-range order
  characteristic of spontaneous symmetry breaking.}\BibitemShut {Stop}%
\bibitem [{Note2()}]{Note2}%
  \BibitemOpen
  \bibinfo {note} {We thank Ehud Altman for this incisive
  question.}\BibitemShut {Stop}%
\bibitem [{\citenamefont {Wilczek}(2012)}]{Wilczek12}%
  \BibitemOpen
  \bibfield  {author} {\bibinfo {author} {\bibfnamefont {F.}~\bibnamefont
  {Wilczek}},\ }\href {\doibase 10.1103/PhysRevLett.109.160401} {\bibfield
  {journal} {\bibinfo  {journal} {Phys. Rev. Lett.}\ }\textbf {\bibinfo
  {volume} {109}},\ \bibinfo {pages} {160401} (\bibinfo {year}
  {2012})}\BibitemShut {NoStop}%
\bibitem [{\citenamefont {Watanabe}\ and\ \citenamefont
  {Oshikawa}(2015)}]{Oshikawa15}%
  \BibitemOpen
  \bibfield  {author} {\bibinfo {author} {\bibfnamefont {H.}~\bibnamefont
  {Watanabe}}\ and\ \bibinfo {author} {\bibfnamefont {M.}~\bibnamefont
  {Oshikawa}},\ }\href {\doibase 10.1103/PhysRevLett.114.251603} {\bibfield
  {journal} {\bibinfo  {journal} {Phys. Rev. Lett.}\ }\textbf {\bibinfo
  {volume} {114}},\ \bibinfo {pages} {251603} (\bibinfo {year}
  {2015})}\BibitemShut {NoStop}%
\bibitem [{Note3()}]{Note3}%
  \BibitemOpen
  \bibinfo {note} {While spatiotemporal order has been discussed for classical
  systems out of equilibrium, e.g. Ref.~\protect \rev@citealpnum {Sancho07}, to
  our knowledge this is the first appearance of such order for quantum
  systems.}\BibitemShut {Stop}%
\bibitem [{\citenamefont {{Chandran}}\ and\ \citenamefont
  {{Sondhi}}(2015)}]{Chandran15}%
  \BibitemOpen
  \bibfield  {author} {\bibinfo {author} {\bibfnamefont {A.}~\bibnamefont
  {{Chandran}}}\ and\ \bibinfo {author} {\bibfnamefont {S.~L.}\ \bibnamefont
  {{Sondhi}}},\ }\href@noop {} {\bibfield  {journal} {\bibinfo  {journal}
  {ArXiv e-prints}\ } (\bibinfo {year} {2015})},\ \Eprint
  {http://arxiv.org/abs/1506.08836} {arXiv:1506.08836 [cond-mat.stat-mech]}
  \BibitemShut {NoStop}%
\bibitem [{\citenamefont {{Else}}\ \emph {et~al.}(2016)\citenamefont {{Else}},
  \citenamefont {{Bauer}},\ and\ \citenamefont {{Nayak}}}]{Q}%
  \BibitemOpen
  \bibfield  {author} {\bibinfo {author} {\bibfnamefont {D.~V.}\ \bibnamefont
  {{Else}}}, \bibinfo {author} {\bibfnamefont {B.}~\bibnamefont {{Bauer}}}, \
  and\ \bibinfo {author} {\bibfnamefont {C.}~\bibnamefont {{Nayak}}},\
  }\href@noop {} {\bibfield  {journal} {\bibinfo  {journal} {ArXiv e-prints}\ }
  (\bibinfo {year} {2016})},\ \Eprint {http://arxiv.org/abs/1603.08001}
  {arXiv:1603.08001 [cond-mat.dis-nn]} \BibitemShut {NoStop}%
\bibitem [{Note4()}]{Note4}%
  \BibitemOpen
  \bibinfo {note} {These symmetries have similar implications, which we do not
  discuss here for brevity.}\BibitemShut {Stop}%
\bibitem [{\citenamefont {Fisher}(1995)}]{Fisher95}%
  \BibitemOpen
  \bibfield  {author} {\bibinfo {author} {\bibfnamefont {D.~S.}\ \bibnamefont
  {Fisher}},\ }\href {\doibase 10.1103/PhysRevB.51.6411} {\bibfield  {journal}
  {\bibinfo  {journal} {Phys. Rev. B}\ }\textbf {\bibinfo {volume} {51}},\
  \bibinfo {pages} {6411} (\bibinfo {year} {1995})}\BibitemShut {NoStop}%
\bibitem [{\citenamefont {Vosk}\ and\ \citenamefont {Altman}(2013)}]{Vosk13}%
  \BibitemOpen
  \bibfield  {author} {\bibinfo {author} {\bibfnamefont {R.}~\bibnamefont
  {Vosk}}\ and\ \bibinfo {author} {\bibfnamefont {E.}~\bibnamefont {Altman}},\
  }\href {\doibase 10.1103/PhysRevLett.110.067204} {\bibfield  {journal}
  {\bibinfo  {journal} {Phys. Rev. Lett.}\ }\textbf {\bibinfo {volume} {110}},\
  \bibinfo {pages} {067204} (\bibinfo {year} {2013})}\BibitemShut {NoStop}%
\bibitem [{Note5()}]{Note5}%
  \BibitemOpen
  \bibinfo {note} {A local (or low depth) unitary is a unitary which can be
  written as $\protect \mathcal {V}=\protect \mathcal {T} e^{-i \DOTSI \intop
  \ilimits@ ^t_0 ds K(s)}$ for some local bounded Hamiltonian $K(t)$, with $t$
  finite in the thermodynamic limit.}\BibitemShut {Stop}%
\bibitem [{\citenamefont {Gopalakrishnan}\ \emph {et~al.}(2015)\citenamefont
  {Gopalakrishnan}, \citenamefont {M\"uller}, \citenamefont {Khemani},
  \citenamefont {Knap}, \citenamefont {Demler},\ and\ \citenamefont
  {Huse}}]{Gopalakrishnan15}%
  \BibitemOpen
  \bibfield  {author} {\bibinfo {author} {\bibfnamefont {S.}~\bibnamefont
  {Gopalakrishnan}}, \bibinfo {author} {\bibfnamefont {M.}~\bibnamefont
  {M\"uller}}, \bibinfo {author} {\bibfnamefont {V.}~\bibnamefont {Khemani}},
  \bibinfo {author} {\bibfnamefont {M.}~\bibnamefont {Knap}}, \bibinfo {author}
  {\bibfnamefont {E.}~\bibnamefont {Demler}}, \ and\ \bibinfo {author}
  {\bibfnamefont {D.~A.}\ \bibnamefont {Huse}},\ }\href {\doibase
  10.1103/PhysRevB.92.104202} {\bibfield  {journal} {\bibinfo  {journal} {Phys.
  Rev. B}\ }\textbf {\bibinfo {volume} {92}},\ \bibinfo {pages} {104202}
  (\bibinfo {year} {2015})}\BibitemShut {NoStop}%
\bibitem [{Note6()}]{Note6}%
  \BibitemOpen
  \bibinfo {note} {In principle we can now identify symmetry defined
  submanifolds based on keeping the emergent symmetries about {\protect \it
  any} fixed point in the $\pi $SG which provides an ``origin independent''
  view of the structure of the phase. The functional form of the perturbed
  unitary $U_{f\lambda }$ \protect \textup {\hbox {\mathsurround \z@ \protect
  \normalfont (\ignorespaces \ref {eq:Ufcanonical}\unskip \@@italiccorr )}} and
  its implications are among the central results of this paper.}\BibitemShut
  {Stop}%
\bibitem [{\citenamefont {Schreiber}\ \emph {et~al.}(2015)\citenamefont
  {Schreiber}, \citenamefont {Hodgman}, \citenamefont {Bordia}, \citenamefont
  {L{\"u}schen}, \citenamefont {Fischer}, \citenamefont {Vosk}, \citenamefont
  {Altman}, \citenamefont {Schneider},\ and\ \citenamefont
  {Bloch}}]{Schreiber2015}%
  \BibitemOpen
  \bibfield  {author} {\bibinfo {author} {\bibfnamefont {M.}~\bibnamefont
  {Schreiber}}, \bibinfo {author} {\bibfnamefont {S.~S.}\ \bibnamefont
  {Hodgman}}, \bibinfo {author} {\bibfnamefont {P.}~\bibnamefont {Bordia}},
  \bibinfo {author} {\bibfnamefont {H.~P.}\ \bibnamefont {L{\"u}schen}},
  \bibinfo {author} {\bibfnamefont {M.~H.}\ \bibnamefont {Fischer}}, \bibinfo
  {author} {\bibfnamefont {R.}~\bibnamefont {Vosk}}, \bibinfo {author}
  {\bibfnamefont {E.}~\bibnamefont {Altman}}, \bibinfo {author} {\bibfnamefont
  {U.}~\bibnamefont {Schneider}}, \ and\ \bibinfo {author} {\bibfnamefont
  {I.}~\bibnamefont {Bloch}},\ }\href {\doibase 10.1126/science.aaa7432}
  {\bibfield  {journal} {\bibinfo  {journal} {Science}\ }\textbf {\bibinfo
  {volume} {349}},\ \bibinfo {pages} {842} (\bibinfo {year} {2015})},\ \Eprint
  {http://arxiv.org/abs/http://science.sciencemag.org/content/349/6250/842.full.pdf}
  {http://science.sciencemag.org/content/349/6250/842.full.pdf} \BibitemShut
  {NoStop}%
\bibitem [{\citenamefont {Kondov}\ \emph {et~al.}(2015)\citenamefont {Kondov},
  \citenamefont {McGehee}, \citenamefont {Xu},\ and\ \citenamefont
  {DeMarco}}]{Kondov2015}%
  \BibitemOpen
  \bibfield  {author} {\bibinfo {author} {\bibfnamefont {S.~S.}\ \bibnamefont
  {Kondov}}, \bibinfo {author} {\bibfnamefont {W.~R.}\ \bibnamefont {McGehee}},
  \bibinfo {author} {\bibfnamefont {W.}~\bibnamefont {Xu}}, \ and\ \bibinfo
  {author} {\bibfnamefont {B.}~\bibnamefont {DeMarco}},\ }\href {\doibase
  10.1103/PhysRevLett.114.083002} {\bibfield  {journal} {\bibinfo  {journal}
  {Phys. Rev. Lett.}\ }\textbf {\bibinfo {volume} {114}},\ \bibinfo {pages}
  {083002} (\bibinfo {year} {2015})}\BibitemShut {NoStop}%
\bibitem [{\citenamefont {Bordia}\ \emph {et~al.}(2016)\citenamefont {Bordia},
  \citenamefont {L\"uschen}, \citenamefont {Hodgman}, \citenamefont
  {Schreiber}, \citenamefont {Bloch},\ and\ \citenamefont
  {Schneider}}]{Bordia2016}%
  \BibitemOpen
  \bibfield  {author} {\bibinfo {author} {\bibfnamefont {P.}~\bibnamefont
  {Bordia}}, \bibinfo {author} {\bibfnamefont {H.~P.}\ \bibnamefont
  {L\"uschen}}, \bibinfo {author} {\bibfnamefont {S.~S.}\ \bibnamefont
  {Hodgman}}, \bibinfo {author} {\bibfnamefont {M.}~\bibnamefont {Schreiber}},
  \bibinfo {author} {\bibfnamefont {I.}~\bibnamefont {Bloch}}, \ and\ \bibinfo
  {author} {\bibfnamefont {U.}~\bibnamefont {Schneider}},\ }\href {\doibase
  10.1103/PhysRevLett.116.140401} {\bibfield  {journal} {\bibinfo  {journal}
  {Phys. Rev. Lett.}\ }\textbf {\bibinfo {volume} {116}},\ \bibinfo {pages}
  {140401} (\bibinfo {year} {2016})}\BibitemShut {NoStop}%
\bibitem [{\citenamefont {{Choi}}\ \emph {et~al.}(2016)\citenamefont {{Choi}},
  \citenamefont {{Hild}}, \citenamefont {{Zeiher}}, \citenamefont
  {{Schau{\ss}}}, \citenamefont {{Rubio-Abadal}}, \citenamefont {{Yefsah}},
  \citenamefont {{Khemani}}, \citenamefont {{Huse}}, \citenamefont {{Bloch}},\
  and\ \citenamefont {{Gross}}}]{Choi2016}%
  \BibitemOpen
  \bibfield  {author} {\bibinfo {author} {\bibfnamefont {J.-y.}\ \bibnamefont
  {{Choi}}}, \bibinfo {author} {\bibfnamefont {S.}~\bibnamefont {{Hild}}},
  \bibinfo {author} {\bibfnamefont {J.}~\bibnamefont {{Zeiher}}}, \bibinfo
  {author} {\bibfnamefont {P.}~\bibnamefont {{Schau{\ss}}}}, \bibinfo {author}
  {\bibfnamefont {A.}~\bibnamefont {{Rubio-Abadal}}}, \bibinfo {author}
  {\bibfnamefont {T.}~\bibnamefont {{Yefsah}}}, \bibinfo {author}
  {\bibfnamefont {V.}~\bibnamefont {{Khemani}}}, \bibinfo {author}
  {\bibfnamefont {D.~A.}\ \bibnamefont {{Huse}}}, \bibinfo {author}
  {\bibfnamefont {I.}~\bibnamefont {{Bloch}}}, \ and\ \bibinfo {author}
  {\bibfnamefont {C.}~\bibnamefont {{Gross}}},\ }\href@noop {} {\bibfield
  {journal} {\bibinfo  {journal} {ArXiv e-prints}\ } (\bibinfo {year}
  {2016})},\ \Eprint {http://arxiv.org/abs/1604.04178} {arXiv:1604.04178
  [cond-mat.quant-gas]} \BibitemShut {NoStop}%
\bibitem [{Note7()}]{Note7}%
  \BibitemOpen
  \bibinfo {note} {We thank Christian Gross for discussions on possible
  experiments.}\BibitemShut {Stop}%
\bibitem [{\citenamefont {{Chandran}}\ \emph {et~al.}(2016)\citenamefont
  {{Chandran}}, \citenamefont {{Pal}}, \citenamefont {{Laumann}},\ and\
  \citenamefont {{Scardicchio}}}]{Chandran20162D}%
  \BibitemOpen
  \bibfield  {author} {\bibinfo {author} {\bibfnamefont {A.}~\bibnamefont
  {{Chandran}}}, \bibinfo {author} {\bibfnamefont {A.}~\bibnamefont {{Pal}}},
  \bibinfo {author} {\bibfnamefont {C.~R.}\ \bibnamefont {{Laumann}}}, \ and\
  \bibinfo {author} {\bibfnamefont {A.}~\bibnamefont {{Scardicchio}}},\
  }\href@noop {} {\bibfield  {journal} {\bibinfo  {journal} {ArXiv e-prints}\ }
  (\bibinfo {year} {2016})},\ \Eprint {http://arxiv.org/abs/1605.00655}
  {arXiv:1605.00655 [cond-mat.dis-nn]} \BibitemShut {NoStop}%
\bibitem [{\citenamefont {{Roy}}\ and\ \citenamefont
  {{Harper}}(2016{\natexlab{b}})}]{Harper16}%
  \BibitemOpen
  \bibfield  {author} {\bibinfo {author} {\bibfnamefont {R.}~\bibnamefont
  {{Roy}}}\ and\ \bibinfo {author} {\bibfnamefont {F.}~\bibnamefont
  {{Harper}}},\ }\href@noop {} {\bibfield  {journal} {\bibinfo  {journal}
  {ArXiv e-prints}\ } (\bibinfo {year} {2016}{\natexlab{b}})},\ \Eprint
  {http://arxiv.org/abs/1602.08089} {arXiv:1602.08089 [cond-mat.str-el]}
  \BibitemShut {NoStop}%
\bibitem [{\citenamefont {{Rehn}}\ \emph {et~al.}(2016)\citenamefont {{Rehn}},
  \citenamefont {{Lazarides}}, \citenamefont {{Pollmann}},\ and\ \citenamefont
  {{Moessner}}}]{Rehn16}%
  \BibitemOpen
  \bibfield  {author} {\bibinfo {author} {\bibfnamefont {J.}~\bibnamefont
  {{Rehn}}}, \bibinfo {author} {\bibfnamefont {A.}~\bibnamefont {{Lazarides}}},
  \bibinfo {author} {\bibfnamefont {F.}~\bibnamefont {{Pollmann}}}, \ and\
  \bibinfo {author} {\bibfnamefont {R.}~\bibnamefont {{Moessner}}},\
  }\href@noop {} {\bibfield  {journal} {\bibinfo  {journal} {ArXiv e-prints}\ }
  (\bibinfo {year} {2016})},\ \Eprint {http://arxiv.org/abs/1603.03054}
  {arXiv:1603.03054 [cond-mat.stat-mech]} \BibitemShut {NoStop}%
\bibitem [{\citenamefont {Kapustin}\ and\ \citenamefont
  {Seiberg}(2014)}]{Seiberg14}%
  \BibitemOpen
  \bibfield  {author} {\bibinfo {author} {\bibfnamefont {A.}~\bibnamefont
  {Kapustin}}\ and\ \bibinfo {author} {\bibfnamefont {N.}~\bibnamefont
  {Seiberg}},\ }\href {\doibase 10.1007/JHEP04(2014)001} {\bibfield  {journal}
  {\bibinfo  {journal} {Journal of High Energy Physics}\ }\textbf {\bibinfo
  {volume} {2014}},\ \bibinfo {pages} {1} (\bibinfo {year} {2014})}\BibitemShut
  {NoStop}%
\bibitem [{\citenamefont {Gaiotto}\ \emph {et~al.}(2015)\citenamefont
  {Gaiotto}, \citenamefont {Kapustin}, \citenamefont {Seiberg},\ and\
  \citenamefont {Willett}}]{Gaiotto2015}%
  \BibitemOpen
  \bibfield  {author} {\bibinfo {author} {\bibfnamefont {D.}~\bibnamefont
  {Gaiotto}}, \bibinfo {author} {\bibfnamefont {A.}~\bibnamefont {Kapustin}},
  \bibinfo {author} {\bibfnamefont {N.}~\bibnamefont {Seiberg}}, \ and\
  \bibinfo {author} {\bibfnamefont {B.}~\bibnamefont {Willett}},\ }\href
  {\doibase 10.1007/JHEP02(2015)172} {\bibfield  {journal} {\bibinfo  {journal}
  {Journal of High Energy Physics}\ }\textbf {\bibinfo {volume} {2015}},\
  \bibinfo {pages} {1} (\bibinfo {year} {2015})}\BibitemShut {NoStop}%
\bibitem [{\citenamefont {Sagu\'es}\ \emph {et~al.}(2007)\citenamefont
  {Sagu\'es}, \citenamefont {Sancho},\ and\ \citenamefont
  {Garc\'{\i}a-Ojalvo}}]{Sancho07}%
  \BibitemOpen
  \bibfield  {author} {\bibinfo {author} {\bibfnamefont {F.}~\bibnamefont
  {Sagu\'es}}, \bibinfo {author} {\bibfnamefont {J.~M.}\ \bibnamefont
  {Sancho}}, \ and\ \bibinfo {author} {\bibfnamefont {J.}~\bibnamefont
  {Garc\'{\i}a-Ojalvo}},\ }\href {\doibase 10.1103/RevModPhys.79.829}
  {\bibfield  {journal} {\bibinfo  {journal} {Rev. Mod. Phys.}\ }\textbf
  {\bibinfo {volume} {79}},\ \bibinfo {pages} {829} (\bibinfo {year}
  {2007})}\BibitemShut {NoStop}%
\end{thebibliography}

%

\begin{appendix}

\section{$\tau^z_{r,\lambda}$ either commutes or anti-commutes with $U_{f\lambda}$}\label{[Z:P]=pm1}
To prove this assertion, we will use only the locality of the $\mathcal{V}_{\lambda},U_{f \lambda}$. First note that we can express a product of any
two $\tau^z_{\lambda}$ operators as a product of l-bits  ${\tau^z_{r,\lambda}\tau^z_{s,\lambda}=\prod_{r}^{s-1} D^\lambda_{r}}$.
This compound operator commutes with with $U_{f,\lambda}$ because the $ D^\lambda_{r}$
do , i.e.,
\be\label{eq:localityandlbits}
U_{f,\lambda}\tau^z_{r,\lambda}\tau^z_{s,\lambda}U_{f,\lambda}^{\dagger}=\tau^z_{r,\lambda}\tau^z_{s,\lambda}
\ee
However note that the unitaries defined as
\begin{align}
\theta_{r} & \equiv \tau^z_{r,\lambda}U_{f,\lambda}\tau^z_{r,\lambda}U_{f,\lambda}^{\dagger}     \label{eq:thetar}\\
\theta_{s} & \equiv \tau^z_{s,\lambda}U_{f,\lambda}\tau^z_{s,\lambda}U_{f,\lambda}^{\dagger}   \label{eq:thetas}
\end{align}
are local to $r,s$ respectively. This follows from two observations.
First $\tau^z_{r,\lambda}$ is local to $r$ because $\mathcal{V}_{\lambda}$
is assumed low depth. Second, $U_{f,\lambda}\tau^z_{r,\lambda}U_{f,\lambda}^{\dagger}$
is local to $r$ because $\tau^z_{r,\lambda}$ is, and $U_{f,\lambda}$
is low depth (being the finite time ordered exponent of a bounded
local Hamiltonian). Plugging \eqnref{eq:thetar} and \eqnref{eq:thetas} into \eqnref{eq:localityandlbits}
gives
\be
U_{f,\lambda}\tau^z_{r,\lambda}\tau^z_{s,\lambda}U_{f,\lambda}^{\dagger}=\tau^z_{r,\lambda}\theta_{r}\theta_{s}^{-1}\tau^z_{s,\lambda}=\tau^z_{r,\lambda}\tau^z_{s,\lambda}
\ee
implying that
\be\label{eq:equality of the thetas}
\theta_{r\lambda}=\theta_{s\lambda}
\ee
despite the fact that $\theta_{r\lambda},\theta_{s\lambda}$ are exponentially
localized to potentially distant sites $r,s$ -- in particular we
could say choose $|r-s|=L/2$ to be of order the system size. The
implication is then that, up to exponentially small corrections in
system size, $\theta_{r,s\lambda}$ are pure phases. The corrections take the form $C e^{- L/\xi}$, where $C,\xi$ do not depend on the system size, and only depends on the details of $\mathcal{V}_\lambda, U_{f,\lambda}$ (such as their depth, which is assumed to be finite). The fact $(\tau^z_{r,\lambda})^2=1$ and $\theta_{r\lambda}$ approximately a pure phase implies $\theta^2_{r\lambda}=1+\epsilon$ where $\epsilon$ is a correction of the form $ce^{-L/\xi}$ and $c=O(1)$. This shows that
\be \label{eq:LR}
\theta_{r\lambda}=\pm1\punc{.}
\ee
to the same degree of a approximation. Supposing we know that $\theta_{r0}=-1$ exactly -- as is the case for the fixed point $\pi$SG model \eqnref{eq:piSG}. If $\mathcal{V}_{\lambda},U_{f\lambda}$ is a continuous family of unitaries it follows by continuity that
$\theta_{r\lambda}=-1$ in the large system limit, for all applicable $\lambda$.

\section{Symmetries and the $\mathcal{V}_\lambda$ unitaries}\label{app:symmetries Vl}
Here we argue that diagonalizing unitaries $\mathcal{V}_\lambda$ for families of unitaries $U_{f\lambda}$ respecting a fixed symmetry (e.g., Ising parity or time reversal) and exhibiting absolutely stable long ranged order, can themselves be chosen to commute with the fixed symmetry.  For concreteness,  focus on a system with an anti-unitary symmetry $\mathcal{T}$ with $\mathcal{T}^2=1$ -- the unitary symmetry case goes through similarly. Thus we consider a family of unitaries $U_{f\lambda}$ obeying  $\mathcal{T}U_{f\lambda}\mathcal{T}U_{f\lambda}=1$, with $U_{f0}$ given by \eqnref{eq:piSG}. Note first that the spectrum of $U_{f0}$ generically has no degeneracies. Assuming the same is true of $U_{f\lambda}$ for now, consider the action of $\mathcal{T}$ on eigenstates. As $\mathcal{T}U_{f\lambda}\mathcal{T}U_{f\lambda}=1$, it follows that $U_{f\lambda}\mathcal{T}\mid\left\{ d\right\} ,p\rangle_{\lambda}=u_{d,p,\lambda}\mathcal{T}\mid\left\{ d\right\} ,p\rangle_{\lambda}$. Hence $\mathcal{T}$ preserves eigenstates of $U_{f\lambda}$. As the eigenstates are non-degenerate it follows that
\be\label{eq:Taction}
\mathcal{T}\mid\left\{ d\right\} ,p\rangle_{\lambda}=e^{i\theta_{d,p}}\mid\left\{ d\right\} ,p\rangle_{\lambda}
\ee
for some state dependent phase $e^{i\theta_{d,p}}$. \eqnref{eq:Taction} immediately implies $ d_{\lambda,r}=\mathcal{T} d_{\lambda,r} \mathcal{T}$ and $ P^{\lambda} =\mathcal{T} P^{\lambda} \mathcal{T}$ which we can rewrite as
\begin{align*}
\mathcal{V}_{\lambda} d_{r} \mathcal{V}_{\lambda}^{-1} & =\mathcal{V}_{\lambda,\mathcal{T}}d_{r}\mathcal{V}_{\lambda,\mathcal{T}}^{-1}\\
\mathcal{V}_{\lambda} P \mathcal{V}_{\lambda}^{-1}&=\mathcal{V}_{\lambda,\mathcal{T}}P\mathcal{V}_{\lambda,\mathcal{T}}^{-1}
\end{align*}
where $\mathcal{V}_{\lambda,\mathcal{T}}\equiv \mathcal{T}\mathcal{V}_{\lambda}\mathcal{T}^{-1}$, and $d_r,P$ are the undressed domain wall and parity operators. The upshot is that the unitary
\be
\mathcal{Q}_{\lambda}\equiv\mathcal{V}_{\lambda}^{-1}\mathcal{V}_{\lambda,\mathcal{T}}
\ee
commutes with the commuting set of operators $\{d_{r}\},P$. As these operators uniquely label a complete basis,  $Q_\lambda$ is completely diagonal in $\{d_r\},P$. In other words it can be expressed as
\be
\mathcal{Q}_{\lambda}=e^{-iq_{\lambda}(d_{r},P)}
\ee
for some real functional $q_{\lambda}$ of the labels. In fact, using
locality arguments similar to those in \secref{[Z:P]=pm1} (and in the appendix to Ref.~\onlinecite{vonKeyserlingkSondhi16a}) we find
\be
\mathcal{Q}_{\lambda}=P^{a}e^{-i s_{\lambda}(\left\{ d\right\} )}
\ee
up to exponentially small corrections in system size, where $a=0,1$, and $s$ is a local functional of domain walls. We can use continuity of $\mathcal{V}_\lambda$ again to argue moreover that $a=0$. Therefore we have shown that $\mathcal{V}_{\lambda,\mathcal{T}}=  \mathcal{V}_{\lambda}\mathcal{Q}_{\lambda}$. We now use this result to construct a new change of basis matrix which is invariant under time reversal. We  define a new change of basis unitary $\mathcal{W}_{\lambda}\equiv\mathcal{V}_{\lambda}e^{-i s_{\lambda}(\left\{ d\right\} )/2}$.  $\mathcal{W}_{\lambda}$  indeed achieves the desired local change of basis, but is also time reversal invariant. We henceforth redefine $\mid\left\{ d\right\} ,p\rangle_{\lambda}\equiv\mathcal{W}_{\lambda}\mid\left\{ d\right\} ,p\rangle$. The operators $d_{\lambda,r},P^{\lambda}$ are unaffected by this change in convention. 
\end{appendix}

\end{document}